# Regression Modeling for Recurrent Events Possibly With an Informative Terminal Event Using R Package reReg

**Sy Han Chiou**
University of Texas at Dallas

**Gongjun Xu**
University of Michigan

**Jun Yan**
University of Connecticut

**Chiung-Yu Huang**
University of California
at San Francisco

**Abstract**

Recurrent event analyses have found a wide range of applications in biomedicine, public health, and engineering, among others, where study subjects may experience a sequence of event of interest during follow-up. The R package **reReg** (Chiou and Huang 2022) offers a comprehensive collection of practical and easy-to-use tools for regression analysis of recurrent events, possibly with the presence of an informative terminal event. The regression framework is a general scale-change model which encompasses the popular Cox-type model, the accelerated rate model, and the accelerated mean model as special cases. Informative censoring is accommodated through a subject-specific frailty without any need for parametric specification. Different regression models are allowed for the recurrent event process and the terminal event. Also included are visualization and simulation tools.

## 1. Introduction

Recurrent event data arise when a study subject can experience a sequence of nonfatal events such as hospital admissions, repeated infection episodes, and tumor recurrences, during follow-up. A simple approach is to perform survival analysis with the first event only; however, this approach discards information on subsequent events and might not properly characterize the covariate effects on the recurrent event process (e.g., Claggett, Pocock, Wei, Pfeffer, McMurray, and Solomon 2018). Thus, approaches that address the sequential feature of the recurrent event times without information loss have attracted considerable attention (e.g., Cook and Lawless 2007; Amorim and Cai 2015; Charles-Nelson, Katsahian, and Schramm 2019).

A few R packages offer nonparametric methods for recurrent events. The `survfit()` function from the **survival** package (Therneau 2021) can compute the Nelson-Aalen estimator (Lawless and Nadeau 1995) of the cumulative intensity function of the recurrent event process, with the standard errors obtained from an infinitesimal Jackknife approach (Efron 1982). This allows users to perform two-sample tests by visually checking if the confidence intervals of the Nelson-Aalen estimates overlap. The **reda** package (Wang, Fu, Chiou, and Yan 2021) also provides an implementation of the Nelson-Aalen estimator via the `mcf()` function, with different options for variance estimation and the pseudo-score test (Cook, Lawless, and Nadeau 1996) for comparing Nelson-Aalen estimate from different groups.

Many regression models have been implemented via R packages. The popular Andersen–Gill (AG) model (Andersen and Gill 1982) can be fit with the `coxph()` function from the **survival** package as a generalization of the Cox relative risk model, with possibly time-varying coefficients. The `cph()` function from the **rms** package (Harrell Jr. 2018) extends the AG model to allow for time-dependent covariates. Both the `coxph()` and `cph()` provide an option to calculate a robust sandwich variance estimator of Lin, Wei, Yang, and Ying (2000) to account for the within-subject dependence. The `tpr()` function from the **tpr** package (Yan 2019) fits a temporal process regression that models the recurrent event process at each time point marginally (Fine, Yan, and Kosorok 2004). The **condGEE** package (Clement 2013) models the recurrent gap time process using methods of generalized estimating equation (Clement and Strawderman 2009). A different approach to account for the association between the recurrent events is to introduce frailty variables in the model. The `rateReg()` function from the **reda** package implements a gamma frailty model (Fu, Luo, and Qu 2016; Ma, Qu, and Fu 2021), whose estimating procedure requires a parametric assumption on the frailty variable and the recurrent event process. All these approaches require the censoring time to be independent of the recurrent event process given the covariates.

Recurrent events can be terminated by an informative censoring or a terminal event. The **frailtypack** package (Rondeau, Mazroui, and González 2012) provides a collection of frailty models to relax the conditional independent censoring assumption and jointly model the recurrent event process and the terminal event. Although these frailty models allow for a positive or negative association between the two outcome process, the validity of the inferences relies on a correct parametric assumption on the frailty variable. For this reason, frailty models that account for informative censoring without imposing parametric assumptions are particularly appealing. Such models can be formulated as, for example, Cox-type models (Wang, Qin, and Chiang 2001; Huang and Wang 2004; Huang, Qin, and Wang 2010) or accelerated mean models (Xu, Chiou, Huang, Wang, and Yan 2017). The regression parameters are estimated without information about the frailty variable. Both types of models are special cases of the generalized scale-change model (Xu, Chiou, Yan, Marr, and Huang 2020). An efficient implementation of the models have been made available in the **reReg** (Chiou and Huang 2022) package, which has not been widely known to users who need them.

The **reReg** package provides a comprehensive collection of practical and easy-to-use tools for exploratory and regression analyses of recurrent events. Features include event plots and Nelson-Aalen estimators to visualize recurrent events and terminal events for each subject with possibly multiple event types through ggplot2 objects (Wickham 2009). The package also provides regression model based on the general semiparametric scale-change model for the recurrent event process (Xu *et al.* 2020) that covers some of the most commonly used forms, including the Cox-type model, the accelerated rate model, and the accelerated mean model, as special cases. In the presence of terminal events, the package allows users to choose to jointly model the recurrent event process with the terminal event via a shared frailty, or to treat the terminal events as nuisances. In the former case, a general scale-change model or a special case of it can be specified for the hazard function of the terminal event. No parametric assumption on the latent frailty variable is needed. Standard errors of the estimates are obtained through efficient resampling procedures, which can run parallel on computers with multicores. These features make the **reReg** package appealing in facilitating recurrent event analyses in many application fields.

The rest of the article is organized as follows. Notation used to describe the underlying and observed recurrent event data structure is introduced in Section 2. The methodological framework and the estimating procedure for modeling the recurrent event process are described in Section 3. The structure of the package is presented in Section 4. Illustrations are provided in Sections 5 and 6. Section 7 concludes with a few remarks.

## 2. Notation for recurrent events

Let $N(t)$ be the number of recurrent events occurring over the interval $[0, t]$ and $D$ be the failure time of interest subjects to right censoring by $C$. Define the composite censoring time $Y = \min(D, C, \tau)$ and the failure event indicator $\Delta = I\{D \leq \min(C, \tau)\}$, where $\tau$ is the maximum follow-up time. We assume the recurrent event process $N(\cdot)$ is observed up to $Y$. Let $X$ be a $p$-dimensional covariate vector. Consider a random sample of $n$ subjects, the observed data are independent and identically distributed (iid) copies of $\{Y_i, \Delta_i, X_i, N_i(t), 0 \leq t \leq Y_i\}$, where the subscript $i$ denotes the index of a subject for $i = 1, \ldots, n$. Let $m_i = N_i(Y_i)$ be the number of recurrent events the $i$th subject experienced before time $Y_i$, then the jump times of $N_i(t)$ give the observed recurrent event times $t_{i1}, \ldots, t_{im_i}$ when $m_i > 0$. Thus, the observed data can also be expressed as iid copies of $\{Y_i, \Delta_i, X_i, m_i, (t_{i1}, \ldots, t_{im_i})\}$. The primary interests in recurrent event data analysis often lie in making inference about the recurrent event process and the failure event and understanding the corresponding covariate effects. The built-in dataset, `simDat`, is a simulated example of a recurrent event data:

```
R> library("reReg")
R> subset(simDat, id %in% c(5, 7, 12))

##    id    t.start      t.stop event status x1         x2
## 25  5  0.0000000  1.1695164     0      1  1  0.6570011
## 30  7  0.0000000  0.4690579     1      0  0 -0.2149894
## 31  7  0.4690579  0.8320469     0      1  0 -0.2149894
## 46 12  0.0000000  3.2220999     1      0  1 -0.2862713
## 47 12  3.2220999 13.5607302     1      0  1 -0.2862713
## 48 12 13.5607302 60.0000000     0      0  1 -0.2862713
```

The `simDat` data set consists of 200 hypothetical subjects, whose event times and covariate information were generated by the `simGSC()` function described in Section 5.6. The data set consists of seven variables. The `id` variable is used to denote the subject identification. The `t.start` and `t.stop` variables mark the start and end of each time interval, respectively. The `event` variable is the recurrent event indicator that indicates whether a recurrent event was observed (`event = 1`) at the end of the time interval. The `status` variable is the failure event indicator ($\Delta$) that indicates whether the recurrent event process was terminated by a terminal event (`status = 1`). Finally, the `x1` variable is a binary covariate and the `x2` variable is a continuous covariate.

The `simDat` data presents the recurrent event in person time, where the beginning of time is zero. Thus, for the $i$th subject, the endpoint of the time intervals, i.e., `t.stop`, represents a recurrent event time ($t_{ij}$) when `event = 1` or a censoring time ($Y_i$) when `event = 0`. For example, in `simDat`, Subject 5 did not experience a recurrent event before the terminal event at 1.170 while Subject 7 experienced a recurrent event at 0.469 before the terminal event at 0.832. On the other hand, Subject 12 experienced two recurrent events at 3.222 and 13.561 before the non-terminal event at 60. We demonstrate the usage of the **reReg** package for data sets in forms similar to `simDat` in Section 5.

## 3. Models and inferences

### 3.1. Nonparametric estimation of rate function

Statistical methods for recurrent event data often involve modeling the recurrent event process via its intensity function or rate function. Define the history of the recurrent event process at time $t$ as $H(t) = \{N(s); 0 \leq s < t\}, t > 0$, the intensity function for $N(t)$ is

$$\lambda\{t|H(t)\} = \lim_{\delta \downarrow 0} \frac{\Pr\{N(t+\delta) - N(t) > 0|H(t)\}}{\delta}, \quad t \in [0, \tau].$$

On the contrary, the rate function for $N(t)$ is unconditional on the history and is defined as

$$\lambda(t) = \lim_{\delta \downarrow 0} \frac{\Pr\{N(t+\delta) - N(t) > 0\}}{\delta}, \quad t \in [0, \tau].$$

The intensity function completely specifies a recurrent event process while the rate function only specifies the population average of the occurrence rate. The latter gives more direct interpretations for explaining covariate effects under weaker assumptions and is generally preferred in practice. The cumulative rate function, defined as

$$\Lambda(t) = \int_0^t \lambda(u) \mathrm{d}u, \quad t \in [0, \tau],$$

is also the expected number of recurrent events occurring in $[0, t]$.

Under independence censoring, a nonparametric estimate of $\Lambda(t)$ known as the Nelson-Aalen estimator is (Lawless and Nadeau 1995)

$$\widehat{\Lambda}(t) = \sum_{i=1}^n \int_0^t \frac{\mathrm{d}N_i(u)}{\sum_{j=1}^n I(Y_i \geq u)}, \tag{1}$$

which is also known as the mean cumulative function (MCF). The MCF presents the average number of recurrent events per subject observed by time $t$ while adjusting for the risk set. In the case when all subjects remain at risk of recurrent events throughout the study, i.e., $n = \sum_{j=1}^{n} I(Y_i \geq t)$, Equation 1 reduces to the cumulative sample mean function introduced in Chapter 1 of Cook and Lawless (2007).

When censoring is not independent of the recurrent event process, nonparametric estimators proposed by Wang *et al.* (2001); Wang and Chiang (2002) that account for informative censoring via a latent frailty variable can be considered. Let $Z$ be a non-negative subject-specific latent frailty variable such that $N(\cdot)$ is conditionally independent of $Y$ given $Z$. The multiplicative intensity model assumes that $N(t)$ is a non-stationary Poisson process with the intensity function $Z\lambda_0(t)$, where the baseline intensity function $\lambda_0(t)$ is a continuous function. Conditioning on $Z$, the independent increments assumption of the Poisson process implies that the rate function is the intensity function. Thus, the rate function and the cumulative rate function can be expressed as $\lambda(t) = \mathsf{E}\{Z\lambda_0(t)\} = \mu_z\lambda_0(t)$ and $\Lambda(t) = \mu_z\Lambda_0(t)$, respectively, where $\mu_z = \mathsf{E}(Z)$ and $\Lambda_0(t) = \int_0^t \lambda_0(u)\mathrm{d}u$. The multiplicative model structure implies that the event occurrence rate is inflated (or deflated) by the frailty variable $Z$.

To estimate $\Lambda(t)$, Wang *et al.* (2001) argued that, given $(Y_i, Z_i, m_i)$, the event times $\{t_{i1}, \ldots, t_{im_i}\}$ are the order statistics of independent copies of random variables with the density function $f(t)/F(Y_i), 0 \leq t \leq Y_i$, where

$$f(t) = \frac{\lambda(t)}{\Lambda(\tau)} = \frac{Z_i\lambda_0(t)}{Z_i\Lambda_0(\tau)} = \frac{\lambda_0(t)}{\Lambda_0(\tau)}, \quad 0 \leq t \leq \tau,$$

and $F(t) = \int_0^t f(u)\mathrm{d}u = \Lambda_0(t)/\Lambda_0(\tau)$. Since the density function $f(t)/F(Y_i)$ is a truncated density function of $f(t)$ with right-truncation on $Y_i$ and does not depends on $Z$, the conditional likelihood based on $\{t_{i1}, \ldots, t_{im_i}\}$ is in the form of

$$L_c \propto \prod_{i=1}^{n} \prod_{k=1}^{m_i} \frac{f(t_{ij})}{F(Y_i)}.$$

The conditional likelihood $L_c$ is maximized at the nonparametric maximum likelihood estimator (NPMLE) (Wang, Jewell, and Tsai 1986)

$$\widehat{F}(t) = \prod_{s_{(\ell)} > t} \left(1 - \frac{d_{(\ell)}}{R_{(\ell)}}\right), \quad (2)$$

where $\{s_{(\ell)}\}$ are the ordered and distinct values of $\{t_{ik}, i = 1, \ldots, n, k = 1, \ldots, m_i\}$, $d_{(\ell)}$ is the number of events occurring at $s_{(\ell)}$, and $R_{(\ell)}$ is the number of events satisfying $t_{ik} \leq s_{(\ell)} \leq Y_i$. It follows from $\mathsf{E}(m_i|Y_i, Z_i) = Z_i\Lambda_0(Y_i)$ that

$$\mathsf{E}\left\{\frac{m_i}{F(Y_i)}\right\} = \mathsf{E}\left[\mathsf{E}\left\{\frac{m_i}{F(Y_i)}\middle| Y_i, Z_i\right\}\right] = \mathsf{E}\left\{\frac{Z_i\Lambda_0(Y_i)}{F(Y_i)}\right\} = \mu_z\Lambda_0(\tau). \quad (3)$$

If $\Lambda_0(\tau) = 1$ is assumed for model identifiability, then $\Lambda_0(t)$ can be estimated by $\widehat{\Lambda}_{0n}(t) = \widehat{F}(Y_i)$ and Equation 3 implies $\widehat{\mu}_Z = n^{-1}\sum_{i=1}^{n} m_i/\widehat{\Lambda}_{0n}(Y_i)$. The estimating procedure does not require any distributional assumption on $Z$, making it more appealing than conventional parametric approaches under informative censoring. This estimator can be plotted by applying the generic function `plot()` to a `reReg` object constructed as an intercept-only-model of (6) as illustrated in Section 5.4.

### 3.2. A joint Cox-type model

The AG model is commonly considered when the interest is in evaluating the multiplicative covariate effect on the rate function of the recurrent event process. The AG model specifies the rate/intensity function of the recurrent event process $N(t)$ as

$$\lambda(t) = \lambda_0(t) e^{X^\top \beta},$$

where $\lambda_0(\cdot)$ is an unspecified baseline rate function, and $\beta$ is a $p$-dimensional regression parameter. The estimate of $\beta$ can be obtained through the partial likelihood approach (Cox 1975) under the Poisson assumption, where the time increments between events are conditionally independent given covariates. When the correlation among recurrent events is not induced by the covariates, Pepe and Cai (1993) and Lin *et al.* (2000) relaxed the Poisson assumption and proposed a robust sandwich variance estimator. Applying the AG model is straightforward as these inference procedures are implemented in major software packages. In the R environment, the AG model with the robust sandwich variance estimator can be fitted by the `coxph()` function from the **survival** package (Therneau 2021) with the subjects' identification specified via the `cluster` option. The same procedure can also be called conveniently with the `reReg()` function as illustrated in Section 5.4.

The AG model and many of its extension require the non-informative censoring assumption, that assumes the recurrent events and the censoring times are independent given covariates. However, the non-informative censoring assumption could be violated when the recurrent event process is terminated by informative dropouts or failure events. Wang *et al.* (2001) extends the AG model to accommodate informative censoring via the use of a frailty variable. Specifically, Wang *et al.* (2001) assumes the Cox-type model

$$\lambda(t) = Z \lambda_0(t) e^{X^\top \beta}, \tag{4}$$

where $Z$ is a non-negative subject-specific latent frailty variable. When covariate effects on both the recurrent event process and the failure time are of interest, Huang and Wang (2004) extends Model 4 to the joint model

$$\begin{cases} \lambda(t) = Z \lambda_0(t) e^{X^\top \beta}, \\ h(t) = Z h_0(t) e^{X^\top \theta}, \quad t \in [0, \tau], \end{cases} \tag{5}$$

where $h_0(\cdot)$ is the baseline hazard function for the failure time, $\theta$ is a $p$-dimensional regression parameter. The frailty variable, $Z$, has a proportional effect on both $\lambda(t)$ and $h(t)$, thus inducing a positive association between the recurrent event process $N(\cdot)$ and the censoring events $(D, C)$. The proportionality assumption implicates a wide range of situations. For example, patients who survived longer tend to have fewer cardiovascular-related hospitalizations (Rogers, Yaroshinsky, Pocock, Stokar, and Pogoda 2016). For model identifiability, we assume $\Lambda_0(\tau) = \int_0^\tau \lambda_0(u) du = 1$ and $\mathsf{E}(Z|X) = \mathsf{E}(Z) = \mu_Z$ for some constant $\mu_Z$. In contrast to many shared-frailty models that require a parametric assumption, the distribution of $Z$ is left unspecified throughout the estimating procedure. The joint model (5) presents a common phenomenon where both $\lambda(t)$ and $h(t)$ can be both inflated (or deflated) by the shared subject-specific frailty $Z$. For example, if the recurrent events are recurrent infections after transplant and the failure event is death, then it is reasonably to presume that patients with higher rate of recurrent events are potentially more vulnerable and are more likely to experience a failure event. Models 4 and 5 are available in **reReg** by calling the `reReg()` function with argument `model = "cox"` and `model = "cox|cox"`, respectively.

The general strategy in the estimating procedure for Model 5 is to first estimate the regression parameter in the rate model and the baseline cumulative rate function, then apply the borrowing-strength method of Huang and Wang (2004) to incorporate recurrent event information to estimate the regression parameters in the hazard model and the baseline cumulative hazard function. In the following, we outline the estimating procedure implemented in **reReg** that is used to estimate the parameters in Model 5. The parameter $\beta$ in the rate model of the Model 5 can be estimated by solving the following estimating equation:

$$\frac{1}{n}\sum_{i=1}^n \bar{X}^\top \left\{ \frac{m_i}{\widehat{\Lambda}_{0n}(Y_i)} - e^{\bar{X}_i^\top \psi} \right\} = 0,$$

where $\bar{X}_i = (\mathbf{1}, X_i)$, $\psi = (\log \mu_Z, \beta^\top)^\top$, $\widehat{\Lambda}_{0n}(\cdot)$ is the NPMLE described in Equation 2, and $\mathbf{1}$ is a vector of 1's. Denote the estimator of $\beta$ by $\widehat{\beta}_n$. Under Model 5, the relationship $\mathsf{E}\{m_i|X_i, Y_i, Z_i\} = Z_i \Lambda_0(Y_i) e^{X_i^\top \beta}$ implies the individual frailty value $Z_i$ can be estimated by

$$\widehat{Z}_i = \frac{m_i}{\widehat{\Lambda}_{0n}(Y_i) e^{X_i^\top \widehat{\beta}_n}}.$$

The score function for $\theta$ can be derived directly from the partial likelihood under the hazard model specified in (5); however, it cannot be evaluated because the frailty variable $Z$ is unobservable. The borrowing-strength method of Huang and Wang (2004) proposed to plug in $\widehat{Z}_i$ for $Z_i$ in the score function. The score functions, with $Z_i$ and $\widehat{Z}_i$, attain the same convergence function and the zero-crossing of the latter serves as an estimator for $\theta$. Following the arguments in Huang and Wang (2004), the estimating equation used to estimate $\theta$ is

$$\frac{1}{n}\sum_{i=1}^n \Delta_i \left\{ X_i - \frac{\sum_{j=1}^n X_j \widehat{Z}_j e^{X_j^\top \theta} I(Y_j \geq Y_i)}{\sum_{j=1}^n \widehat{Z}_j e^{X_j^\top \theta} I(Y_j \geq Y_i)} \right\} = 0.$$

Given the estimates, $\widehat{\beta}_n$ and $\widehat{\theta}_n$, the baseline cumulative hazard function can be estimated by

$$\widehat{H}_0(t; \widehat{\theta}_n, \widehat{Z}) = \sum_{i=1}^n \frac{\Delta_i I(Y_i \leq t)}{\sum_{j=1}^n \widehat{Z}_j e^{X_j^\top \widehat{\theta}_n} I(Y_i \leq Y_j)}.$$

Detailed derivations of estimating equations as well as the asymptotic properties of the estimators can be found in Huang and Wang (2004).

The nonparametric bootstrap method for clustered data is adopted to estimate the standard errors of the estimators. The bootstrap samples are formed by resampling the subjects with replacement of the same size from the original data. The above estimating procedures are then applied to each bootstrap sample to provide one draw of the bootstrap estimate. With a large number of replicates, the asymptotic variance matrices are estimated by the sample variance of the bootstrap estimates. The borrowing-strength approach yields stable performance in most of our simulation studies, but numerical issues occasionally arise when the majority of $\widehat{Z}_i$'s are zero or when extremely large values are observed in $\widehat{Z}_i$. In the latter case, extremely large values of $\widehat{Z}_i$ are more likely to occur when evaluating $\widehat{\Lambda}_{0n}(\cdot)$ at small $Y_i$'s where the number of events satisfying $s_{(\ell)} \leq Y_i$ is small. We observe that such numerical issues are more frequent in the bootstrap procedure. In an attempt to enhance numerical stability, we provide an option for a heuristic adjustment (Wang, Ma, and Yan 2013; Chiou, Xu, Yan, and Huang 2018) that replaces $\widehat{Z}_i$ with

$$\widehat{Z}_i = \frac{m_i + \epsilon}{\widehat{\Lambda}_{0n}(Y_i)e^{X_i^\top \widehat{\beta}_n} + \epsilon},$$

for a small constant $\epsilon < \min_{1 \leq i \leq n}\{\widehat{\Lambda}_{0n}(Y_i)e^{X_i^\top \widehat{\beta}_n}\}$. A warning message will be issued if a convergence issue is detected in `reReg()`. When a subset of bootstrap estimates do not converge, the asymptotic variance matrices are estimated by the sample variance of the converged bootstrap estimates.

### 3.3. A generalized joint frailty scale-change model

The Cox-type models in Equations 4 and 5 assumes that the covariates have proportionate effects on the rate function of the recurrent event process and the hazard function of the failure event over time. When the proportionality assumption is not met or when the Cox-type models cannot properly characterize the covariate effects the **reReg** package provides a general class of semiparametric scale-change models that includes the Cox-type models as special cases. In the same spirit of the joint model in Equation 5, we consider the generalized joint frailty scale-change model of the form

$$\begin{cases} \lambda(t) = Z\lambda_0(te^{X^\top\alpha})e^{X^\top\beta}, \\ h(t) = Zh_0(te^{X^\top\eta})e^{X^\top\theta}, \quad t \in [0,\tau], \end{cases} \quad (6)$$

where the $p$-dimensional regression coefficients $(\alpha, \eta)$ and $(\beta, \theta)$ correspond to the shape and the size parameters, respectively. We impose the same model identifiability assumption as in Model 5 and assume $\Lambda_0(\tau) = \int_0^\tau \lambda_0(u)\mathrm{d}u = 1$ and $\mathsf{E}(Z|X) = \mathsf{E}(Z) = \mu_Z$ for some constant $\mu_Z$, without making a full parametric assumption on $Z$. We further assume $\lambda_0(t)$ and $h_0(t)$ are not in the Weibull class, i.e., $\lambda_0(t) \not\propto t^q$ and $h_0(t) \not\propto t^r$ for some $q$ and $r$. The Weibull class assumption is also assumed for models that focus only on the rate model or the hazard model of the joint model in Equation 6 (e.g., Chen and Jewell 2001; Sun and Su 2008; Xu *et al.* 2020).

The flexible formulation of (6) includes several popular semiparametric models as special cases, some of which have been recognized by different authors. For example, in the absence of the shape parameters, i.e., $\alpha = \eta = 0$, Model 6 reduces to the joint frailty Cox-type model in (5). On the other hand, in the absence of the size parameters, i.e., $\beta = \theta = 0$, Model 6 reduces to the joint frailty accelerated rate model. When the shape and size parameters are assumed equal, i.e., $\alpha = \beta$ and $\eta = \theta$, Model 6 reduces to the joint frailty accelerated mean model considered in Xu *et al.* (2017). Setting $\eta = \theta = 0$, Model 6 reduces to the generalized scale-change model considered in Xu *et al.* (2020), which also includes different types of rate models as special cases. The Weibull class assumption excludes the scenario when the three submodels (the Cox-type model, the accelerated rate model, and the accelerated mean model) coincide, which results in the identifiability between the shape and the size parameters. On the other hand, the Weibull class assumption is not needed for the submodels because they only involve one type of regression coefficient. A possible approach for checking the Weibull class assumption is described in Xu *et al.* (2020), which is motivated by the fact that, under the Weibull model, the proposed model reduces to the Cox-type model of Wang *et al.* (2001), and $\log \Lambda_0(t)$ is linear in $\log(t)$.

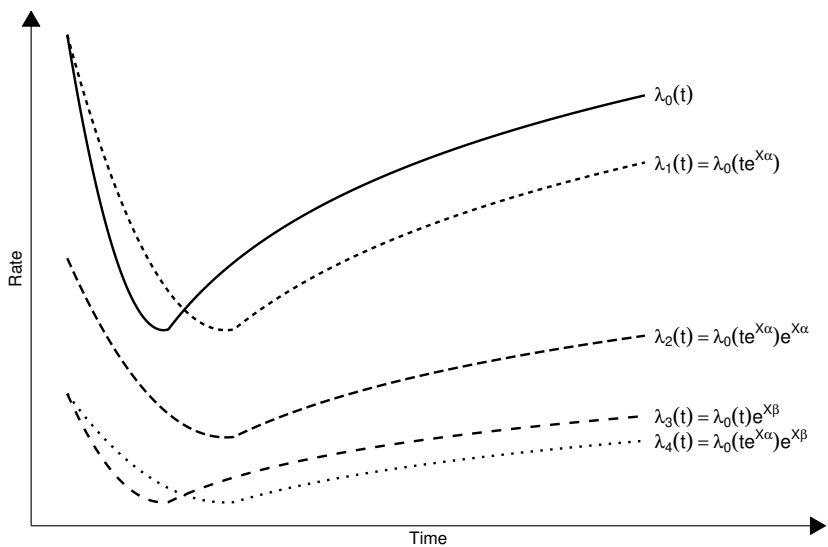

Figure 1: Illustration of the covariate effects on the rate function in Model 6 when $\alpha < 0$, $\beta < 0$, and $X > 0$. An accelerated rate model modifies the time-scale of $\lambda_0(t)$ to $\lambda_1(t)$. An accelerated mean model modifies $\lambda_0(t)$ to $\lambda_2(t)$ in a way that the cumulative mean function of $\lambda_2(t)$ is a time-scale change of that of $\lambda_0(t)$. A Cox-type model modifies the scale of $\lambda_0(t)$ to $\lambda_3(t)$. The proposed model modifies the time-scale and the magnitude of $\lambda_0(t)$ simultaneously to $\lambda_4(t)$, either through $\lambda_0(t) \to \lambda_1(t) \to \lambda_4(t)$ or through $\lambda_0(t) \to \lambda_3(t) \to \lambda_4(t)$.

The joint model (6) allows two types of covariate effects: the shape parameters ($\alpha$ and $\eta$) impose a scale-change effect that alters the time scale and the size parameters ($\beta$ and $\theta$) impose a multiplicative effect that modifies the risk and hazard. In addition, the differences of the shape and size parameters, $\beta - \alpha$ and $\theta - \eta$, modify the cumulative risk and hazard, respectively. Figure 1 illustrates the different covariate effects on the rate function where $\lambda_i(t)$, $i = 0, \ldots, 4$, are hypothetical rate functions under the different model specifications of (6). A one-unit increase in the covariate would change $\lambda_0(t)$ to $\lambda_1(t)$ by modifying the time scale by a factor of $e^\alpha$ and then change $\lambda_1(t)$ to $\lambda_4(t)$ by modifying the magnitude by a factor of $e^\beta$; or change $\lambda_0(t)$ to $\lambda_3(t)$ by modifying the magnitude by a factor of $e^\beta$ and then change $\lambda_3(t)$ to $\lambda_4(t)$ by modifying the time scale by a factor of $e^\alpha$. If $X$ is a treatment indicator, then $e^\beta$ characterizes the risk ratio between the treatment group ($X = 1$) at time $t$ and the control group ($X = 0$) at time $te^\alpha$. When $\alpha = \beta$, the combined changes in time scale and in magnitude are in such a way that the resulting cumulative mean function has a time scale modification. Similar arguments can be made when interpreting $\eta$ and $\theta$ on the hazard function.

The estimating procedure for the generalized joint frailty scale-change model is motivated by those used in Huang and Wang (2004) and Xu *et al.* (2020). Let $a$ be a $p$-dimensional vector and define the transformations $t_{ij}^*(a) = t_{ij} e^{X_i^\top a}$ and $Y_i^*(a) = Y_i e^{X_i^\top a}$. Following the estimating procedure proposed in Xu *et al.* (2020), the shape parameter, $\alpha$, can be estimated by solving the following estimating equation

$$S_{1n}(\alpha) = \frac{1}{n}\sum_{i=1}^{n}\sum_{k=1}^{m_i}\phi_i(\alpha)\left[X_i - \frac{\sum_{j=1}^{n}\sum_{l=1}^{m_j}X_j I\{t_{jl}^*(\alpha) \leq t_{ik}^*(\alpha) \leq Y_j^*(\alpha)\}}{\sum_{j=1}^{n}\sum_{l=1}^{m_j}I\{t_{jl}^*(\alpha) \leq t_{ik}^*(\alpha) \leq Y_j^*(\alpha)\}}\right] = 0,$$

where $\phi_i(\alpha)$, $i = 1, \ldots, n$, is a weight function. Some common choices of $\phi(\alpha)$ are 1 and $n^{-1}\sum_{j=1}^{n}\sum_{l=1}^{m_j}I\{t_{jl}^*(\alpha) \leq t_{ik}^*(\alpha) \leq Y_j^*(\alpha)\}$ corresponding to the log-rank weight and the Gehan weight, respectively. Both the log-rank weight and the Gehan weight are implemented in `reReg()`.

With $\widehat{\alpha}_n$ as the estimator of $\alpha$, the baseline cumulative rate function can be estimated by

$$\widehat{\Lambda}_{0n}(t; \widehat{\alpha}_n) = \exp\left[-\sum_{i=1}^{n}\sum_{k=1}^{m_i}\frac{I\{t_{ik}^*(\widehat{\alpha}_n) \geq t\}}{\sum_{j=1}^{n}\sum_{l=1}^{m_j}I\{t_{jl}^*(\widehat{\alpha}_n) \leq t_{ik}^*(\widehat{\alpha}_n) \leq Y_j^*(\widehat{\alpha}_n)\}}\right].$$

Let $\bar{X}_i = (\mathbf{1}, X_i)$, $\psi = (\log \mu_Z, \gamma^\top)^\top$, and $\gamma = \beta - \alpha$. Given $\widehat{\alpha}_n$ and $\widehat{\Lambda}_{0n}(\cdot)$, $\gamma$ can be estimated by solving the following estimating equation for $\psi$:

$$S_{2n}(\psi) = \frac{1}{n}\sum_{i=1}^{n}\bar{X}_i^\top\left[\frac{m_i}{\widehat{\Lambda}_{0n}\{Y_i^*(\widehat{\alpha}_n)\}} - e^{\bar{X}_i^\top\psi}\right] = 0.$$

Then $\beta$ can be estimated by $\widehat{\beta}_n = \widehat{\alpha}_n + \widehat{\gamma}_n$, where $\widehat{\psi}_n = (\log \widehat{\mu}_Z, \widehat{\gamma}_n^\top)^\top$ is the root of $S_{2n}(\psi) = 0$. The estimating function, $S_{2n}(\psi)$, is monotone and continuously differentiable with respect to $\psi$, and its root can be easily obtained. The detailed derivation of the estimating procedure and the asymptotic properties for $\alpha$, $\beta$, and $\Lambda_0(\cdot)$ can be found in Xu *et al.* (2020).

With the estimated parameters from the rate model, the aforementioned borrowing-strength procedure can be extended to estimate the parameters in the hazard model. Conditioning on $\{X_i, Y_i^*(\alpha), Z_i\}$, the expected value of $m_i$ under Model 6 is

$$\mathsf{E}\{m_i | X_i, Y_i^*(\alpha), Z_i\} = Z_i \Lambda_0\{Y_i^*(\alpha)\}e^{X_i^\top(\beta-\alpha)},$$

which implies the individual frailty value $Z_i$ can be estimated by

$$\widehat{Z}_i = \frac{m_i}{\widehat{\Lambda}_{0n}\{Y_i^*(\widehat{\alpha}_n)\}e^{X_i^\top(\widehat{\beta}_n - \widehat{\alpha}_n)}}.$$

The following estimating equations used to estimate $\eta$ and $\theta$ in the hazard model are generalized from these proposed in Chen and Jewell (2001) to accommodate informative censoring through $Z$:

$$S_{3n}(\eta, \theta) = \sum_{i=1}^{n}\Delta_i\varphi_i(\eta, \theta)\left[X_i - \frac{\sum_{j=1}^{n}\widehat{Z}_j X_j e^{X_j^\top(\theta-\eta)}I\{Y_j^*(\eta) \geq Y_i^*(\eta)\}}{\sum_{j=1}^{n}\widehat{Z}_j e^{X_j^\top(\theta-\eta)}I\{Y_j^*(\eta) \geq Y_i^*(\eta)\}}\right] = 0,$$

$$S_{4n}(\eta, \theta) = \sum_{i=1}^{n}\Delta_i\varphi_i(\eta, \theta)\left[X_i Y_i^*(\eta) - \frac{\sum_{j=1}^{n}\widehat{Z}_j X_j Y_j^*(\eta)e^{X_j^\top(\theta-\eta)}I\{Y_j^*(\eta) \geq Y_i^*(\eta)\}}{\sum_{j=1}^{n}\widehat{Z}_j e^{X_j^\top(\theta-\eta)}I\{Y_j^*(\eta) \geq Y_i^*(\eta)\}}\right] = 0,$$

where $\varphi_i(\eta, \theta)$, $i = 1, \ldots, n$, is a weight function that plays a similar role as $\phi(\alpha)$ in $S_{1n}(\alpha)$. The weight function $\varphi_i(\eta, \theta) = 1$ and $\varphi_i(\eta, \theta) = \sum_{j=1}^{n}\widehat{Z}_j e^{X_j^\top(\theta-\eta)}I\{Y_j^*(\eta) \geq Y_i^*(\eta)\}$ correspond to the log-rank weight and the Gehan weight, respectively. Given the estimates, $\widehat{\eta}_n$ and $\widehat{\theta}_n$, that satisfy $S_{3n}(\widehat{\eta}_n, \widehat{\theta}_n) = 0$ and $S_{4n}(\widehat{\eta}_n, \widehat{\theta}_n) = 0$, the baseline cumulative hazard function can be estimated with

$$\widehat{H}_0(t; \widehat{\eta}_n, \widehat{\theta}_n) = \sum_{i=1}^{n} \frac{\Delta_i I\{Y_i^*(\widehat{\eta}_n) \leq t\}}{\sum_{j=1}^{n} \widehat{Z}_j e^{X_j^\top (\widehat{\theta}_n - \widehat{\eta}_n)} I\{Y_i^*(\widehat{\eta}_n) \leq Y_j^*(\widehat{\eta}_n)\}}.$$

Several equation solvers are available in the package to find the root of the estimating equations. The default solver is the derivative-free Barzilai-Borwein spectral method (Barzilai and Borwein 1988). This method is an iterative procedure that updates the solution guided by a scalar spectral steplength and a linear search direction. We found that choosing good initial values for $(\eta, \theta)$ improves the convergence performance when solving for $S_{3n}(\eta, \theta) = 0$ and $S_{4n}(\eta, \theta) = 0$, while it is less sensitive to the choice of the initial value when solving for $S_{1n}(\alpha) = 0$ and $S_{2n}(\psi) = 0$. The heuristic adjustment for $\widehat{Z}_i$ is

$$\widehat{Z}_i = \frac{m_i + \epsilon}{\widehat{\Lambda}_{0n}\{Y_i^*(\widehat{\alpha}_n)\} e^{X_i^\top(\widehat{\beta}_n - \widehat{\alpha}_n)} + \epsilon}$$

for some small constant $\epsilon < \min_{1 \leq i \leq n} \left[\widehat{\Lambda}_{0n}\{Y_i^*(\widehat{\alpha}_n)\} e^{X_i^\top(\widehat{\beta}_n - \widehat{\alpha}_n)}\right]$.

In addition to the nonparametric bootstrap approach, an efficient resampling-based sandwich estimator proposed in Xu *et al.* (2020) is implemented to estimate the asymptotic variance for $\alpha$ and $\beta$ in Model 6. The estimating equation to obtain the regression parameters in the special cases of Model 6 can be constructed by modifying the above estimating equations. For example, the estimating equation used in estimating $\beta$ in the Model 4 and the rate model in Model 5 can be constructed by setting $\alpha = 0$ in $S_{2n}(\psi) = 0$. When the rate model is an accelerated rate model ($\beta = 0$), the estimating procedure only consists of solving $S_{1n}(\alpha) = 0$. On the other hand, when the rate model is an accelerated mean model ($\alpha = \beta$), the estimating equation is in the form of $S_{2n}(\alpha) = 0$, i.e.,

$$\frac{1}{n} \sum_{i=1}^{n} \bar{X}_i^\top \left[\frac{m_i}{\widehat{\Lambda}_{0n}\{Y_i^*(\widehat{\alpha}_n)\}} - \widehat{\mu}_Z\right] = 0,$$

where $\widehat{\mu}_Z$ is the sample mean of $m_i \widehat{\Lambda}_{0n}^{-1}\{Y_i^*(\widehat{\alpha}_n)\}$ for $i = 1, \ldots, n$. Similarly, the estimating equation for the different form of the hazard model can be constructed by modifying $S_{3n}(\eta, \theta)$ with the corresponding relationship between $\eta$ and $\theta$. For example, when the hazard function is in the Cox-type form as in the Model 5, the estimating equation used to estimate $\theta$ can be constructed by $S_{3n}(0, \theta)$. On the other hand, when the hazard function is an accelerated rate model ($\theta = 0$) or an accelerated failure time model ($\eta = \theta$), the estimating equations can be constructed by using $S_{3n}(\eta, 0)$ or $S_{3n}(\eta, \eta)$, respectively. The estimation of the individual frailty and the baseline functions in the special cases of Model 6 can be modified similarly. Since the estimation of the regression parameters in the hazard model only depends the parameters in the rate model through the estimate of the individual frailty, the above estimating procedure can be adopted to scenarios where the rate model and the hazard model do not have the same form.

The structure of Model 6 facilitates model selection among the submodels via hypothesis testing of model parameters. For example, as described in Xu *et al.* (2020), the Cox-type proportional assumption in the rate model can be tested through $H_0 : \alpha = 0$ versus $H_1 : \alpha \neq 0$ under Model 6. In this case, the test statistic can be constructed with $T_{\text{cox}} = \widehat{\alpha}_n^\top \widehat{\Sigma}(\widehat{\alpha}_n)^{-1} \widehat{\alpha}_n^\top$, where $\Sigma(\widehat{\alpha}_n)^{-1}$ is the estimated covariance matrix of $n^{1/2}(\widehat{\alpha}_n - \alpha)$. Xu *et al.* (2020) has shown that $T_{\text{cox}}$ converges weakly to a Chi-square distribution with $p$ degrees of freedom. Similarly, test statistics $T_{\text{am}} = \widehat{\gamma}_n^\top \widehat{\Sigma}(\widehat{\gamma}_n)^{-1}\widehat{\gamma}_n$ and $T_{\text{ar}} = \widehat{\beta}_n^\top \widehat{\Sigma}(\widehat{\beta}_n)^{-1}\widehat{\beta}_n$ can be used to test the null hypotheses $H_0 : \gamma = 0$ and $H_0 : \beta = 0$, respectively.

## 4. Package structure

The **reReg** package provides useful functions for data exploration, nonparametric estimation, regression analysis, and recurrent event data simulation. Most functions in the **reReg** package are built around recurrent event objects, which are created by the `Recur()` function imported from the **reda** package. The recurrent event object plays a similar role as the survival object created by the `Surv()` function in the **survival** package as both are used as a response variable in a model formula for the respective packages. However, the recurrent event object is designed to track the study subjects' recurrent times, in addition to failure time and censoring status.

The main function for regression modeling in recurrent event analysis is `reReg()`. The implementation of `reReg()` is based on a series of recently developed methods that accommodate informative censoring via a unspecified frailty variable and allows users to either treat the terminal time as a nuisance or incorporate it in a joint model framework. Once a regression model is fitted with `reReg()`, the S3 methods, including `coef()`, `vcov()`, and `summary()` can be applied to a `reReg` object to easily extract relevant regression results. The S3 `plot` method can be applied to obtain the baseline cumulative rate/hazard function and the predicted cumulative rate/hazard functions function. The plot method utilizes the **ggplot2** plotting environment (Wickham 2009) for visualization to allow extensive flexibility and customization. When an intercept-only model is specified with `reReg()`, the proposed method provides nonparametric estimation based on the NPMLE of Wang *et al.* (2001). Furthermore, the `reReg()` function includes some of the common recurrent event models such as the AG model as alternative options.

Besides fitting regression models, simple descriptive statistics and event plot of the recurrent event process can be obtained by directly apply the S3 methods `summary()` and `plot()` to the recurrent event object, created by `Recur()`. For a more sophisticated event plot, the recurrent event object can then be passed to the `plotEvents()` function in a model formula. For convenience, and as a counterpart to the NPMLE of Wang *et al.* (2001) in the absence of informative censoring, the function `mcf()` is imported from the **reda** package to compute the MCF estimates. The **reReg** package allows users to simulate recurrent event data from the generalized frailty joint scale-change model in Equation 6 via the `simGSC()` function. The `simGSC()` function provides a great flexibility in the specification of the baseline rate/function functions, regression parameters, covariate information, censoring distribution, and frailty distribution. Table 1 summarizes the functions mentioned above as well as their compatible S3 methods. These functions are then illustrated with simulated data in Section 5 and are applied to real data in Section 6.

## 5. Illustrations

### 5.1. Creating a recurrent event object

| Functions | Purpose | Compatible generic functions |
|---|---|---|
| *Functions originated from* **reReg** | | |
| reReg() | Fits regression models under the informative censoring assumption | coef(), plot(), summary(), vcov() |
| plotEvents() | Creates event plots based on model formula | |
| simGSC() | Generates recurrent event data | summary() |
| *Functions used in* **reReg** *imported from* **reda** | | |
| Recur() | Creates recurrent event objects | plot(), summary() |
| %2% | Specifies time segments in Recur | |
| mcf() | Creates mean cumulative functions under non-informative censoring | plot() |

Table 1: An overview of the main functions in **reReg**.

The `Recur()` function imported from the **reda** package (Wang *et al.* 2021) creates recurrent event objects needed for **reReg**. The function interface of `Recur()` is

```
R> args(reda::Recur)

## function (time, id, event, terminal, origin, check = c("hard",
##     "soft", "none"), ...)
## NULL
```

The event times are specified via the `time` argument as a vector that represents the time of recurrent events and censoring, or as a list of time intervals that contains the starting time and the ending time of the interval. In the latter, the intervals are assumed to be open on the left and closed on the right, where the right end points are the time of recurrent events and censoring. When the events are recorded in patient times, where each patient enters the study at time zero as in the `simDat` data, the `time` argument can be specified with `time = t.stop` or with `time = t.start %to% t.stop`. The infix operator `%to%` is used to create a list of two elements containing the endpoints of the time intervals. When the patients have different starting time, the different starting time can be specified via the `origin` argument. Alongside the event times, optional arguments including `id`, `event`, and `terminal`, are used to represent the subject's identification, the recurrent event indicator, and the failure event indicator, respectively. In addition to numeric values, `Date` and `difftime` are also allowed. Finally, the `check` argument specifies how to perform validation checks.

The following code creates a `Recur` object from the `simDat` data.

```
R> (reObj <- with(simDat, Recur(t.start %to% t.stop, id, event, status)))

## [1] 1: (0.0000, 0.1817], (0.1817, 0.8400], ..., (2.5367, 2.8402*]
## [2] 2: (0.0000, 1.0852], (1.0852, 7.5907], ..., (37.5591, 60.0000+]
## [3] 3: (0.0000, 0.8123], (0.8123, 4.0266], ..., (9.3434, 10.2991+]
## [4] 4: (0.0000, 0.3582], (0.3582, 1.4426], ..., (39.2478, 60.0000+]
```

```
## [5] 5: (0.0000, 1.1695*]
##  [ reached getOption("max.print") -- omitted 195 entries ]
```

The `Recur` object is an `S4` class object and the `show()` method for the `Recur` object presents recurrent events in intervals. The endpoint of the last interval within each subject indicates the failure time and its censoring status; `*` and `+` indicate whether the recurrent process is censored by a terminal event or non-terminal event, respectively. For concise print, the maximum number of intervals printed per subject is limited to three controlled by `options("reda.Recur.maxPrint")`. For example, in `simDat`, subjects 1 to 4 experienced more than three events, of which only the first two and the last intervals are displayed. The generic method function `summary` can be applied to an `Recur` to produce simple descriptive statistics.

```
R> summary(reObj)

## Call:
## Recur(time = t.start %to% t.stop, id = id, event = event, terminal = status)
##
## Sample size:                                      200
## Number of recurrent event observed:               674
## Average number of recurrent event per subject:    3.37
## Proportion of subjects with a terminal event:     0.59
## Median follow-up time:                            4.735
## Median time-to-terminal event:                    6.975
```

### 5.2. Creating event plots

An easy way to glance at recurrent event data is by plotting the event plots, which can be created by several approaches in the **reReg** package. One way to create a unstratified event plot is to apply the generic function `plot()` function to `Recur` objects. The `plot` method for `Recur` objects internally calls the `plotEvent()` function to create event plots, which is created in the `ggplot2` environment (Wickham 2009) to allow extensive flexibility and customization. The synopsis of `plotEvents()` is:

```
R> args(plotEvents)

## function (formula, data, result = c("increasing", "decreasing"),
##     "asis"), calendarTime = FALSE, control = list(), ...)
## NULL
```

The required argument for `plotEvent()` is `formula`, which contains an `Recur` object and categorical covariates separated by `~`, where the categorical covariates are used to stratify the event plots. Plotting the `Recur` object by `plot()` is only available in **reReg** and is equivalent to calling the `plotEvent()` function with an intercept-only-model. The `data` argument is an optional data frame contains the variables occurring in the `formula`. The `result` argument is an optional character string indicating whether to arrange the subjects by their follow-up times in an ascending order (default), descending order, or plot as-is. The `calendarTime` argument is a logical value indicating whether to create the event plot in calendar time or patient time. When plotting in calendar time, the `result` argument sorts the subjects by the chronological order of terminal events. The `plotEvent()` also includes convenient graphical parameters for plot annotation that can be specified in the `control` argument. The full list of these parameters with their default values are displayed below.

```
R> args(reReg::plotEvents.control)

## function (xlab = NULL, ylab = NULL, main = NULL, terminal.name = NULL,
##     recurrent.name = NULL, recurrent.type = NULL, legend.position = NULL,
##     base_size = 12, cex = NULL, alpha = 0.7, width = NULL, bar.color = NULL,
##     recurrent.color = NULL, recurrent.shape = NULL, recurrent.stroke = NULL,
##     terminal.color = NULL, terminal.shape = NULL, terminal.stroke = NULL,
##     not.terminal.color = NULL, not.terminal.shape = NULL)
## NULL
```

The parameters `xlab`, `ylab`, and `main` are adopted from the base R **graphics** package for modifying labels to the x-axis, the y-axis, and the main title, respectively. The parameters `terminal.name` and `recurrent.name` are character labels displayed in the legend. The parameter `recurrent.type` specifies the recurrent event types to be displayed in the legend when more than one recurrent event types is specified in the `Recur` object. The parameters `legend.position` and `base_size` are adopted from the **ggplot2** theme to control the location of the legend box and the base font size in `pts`, respectively. The remaining parameters control the event plot appearance, including the size (`cex`) and transparency (`alpha`) of the symbols, the width (`width`) and color (`bar.color`) of the event bars, and the color, shape, and stroke of the recurrent event, terminal event, and non-terminal event symbols (`recurrent.color`, `recurrent.shape`, `recurrent.stroke`, etc.).

The following code demonstrates the `plot` method for the `Recur` object, with the resulting unstratified event plot displayed in Figure 2a.

```
R> plot(reObj)
```

The above `plot` method is an alias for the following `plotEvents()` call.

```
R> plotEvents(Recur(t.start %to% t.stop, id, event, status) ~ 1,
+     data = simDat)
```

In the event plot, the gray horizontal lines represent each subjects' follow-up times, ● marks the recurrent events, and ▲ marks terminal events. With the default setting, the event plot is arranged so that subjects with longer follow-up times are presented on the top. Stratifying the event plot could provide new insights into the different covariate effects. For example, an event plot stratified by the binary variable `x1` is presented in Figure 2b, which is created by specifying a formula with the `Recur` object as the response variable and `x1` as a covariate in `plotEvents()`.

```
R> plotEvents(Recur(t.start %to% t.stop, id, event, status) ~ x1,
+    data = simDat)
```

In this case, subjects with `x1 = 1` seems more likely to experience recurrent events and terminal event sooner. The `plotEvents()` can accommodate multiple recurrent types specified via the `event` argument in the `Recur` object. To illustrate this feature, we perturb the event indicator so that the value of `event` indicates the index of the different recurrent event. The following command is used to create the stratified event plot with different recurrent event types presented in Figure 2c.

```
R> simDat$event2 <- with(simDat, ifelse(t.stop > 10 & event > 0, 2, event))
R> plotEvents(Recur(t.start %to% t.stop, id, event2, status) ~ x1,
+    data = simDat)
```

The different types of recurrent events are marked in different colors in Figure 2c. The `plotEvents()` can also create event plots in calendar time by setting `calendarTime = TRUE`. For illustration, we create a new simulated data based on `simDat` with time intervals shift proportionally to `x2`. We further convert the time intervals to a `Date` class. The construction of the new data and the stratified event plot in Figure 2d is included in the following

```
R> simDat2 <- simDat
R> simDat2$t.start <- as.Date(simDat2$t.start + simDat2$x2 * 5,
+    origin = "20-01-01")
R> simDat2$t.stop <- as.Date(simDat2$t.stop + simDat2$x2 * 5,
+    origin = "20-01-01")
R> plotEvents(Recur(t.start %to% t.stop, id, event, status) ~ x1,
+    data = simDat2, calendarTime = TRUE)
```

In this case, subjects with later censoring times are presented on the top and the date string are printed on the axis label of the event plot.

### 5.3. Plotting mean cumulative functions

The MCF is useful in describing and comparing the average number of events of an individual and between groups. Thus, it provides additional insights into the longitudinal patterns of the recurrent process. To create the NPMLE of Wang *et al.* (1986) under the informative censoring assumption, one can fit an intercept-only model with `reReg()`, then applying the generic function `plot()` to the `reReg` object. The following code illustrates this feature with output displayed in Figure 3a.

```
R> mcf1 <- reReg(Recur(t.start %to% t.stop, id, event, status) ~ 1,
+    data = simDat, B = 200)
R> plot(mcf1)
```

The 95% confidence interval is computed based on the non-parametric bootstrap approach with 200 bootstrap replicates as specified by the argument `B = 200`. Separate MCFs can be created and plotted for each level of a factor variable using the `subset` option in `reReg()`. The `basebind()` function can then be applied to combine the MCFs into one plot to allow easy visual comparison. The synopsis for `basebind()` is shown below.

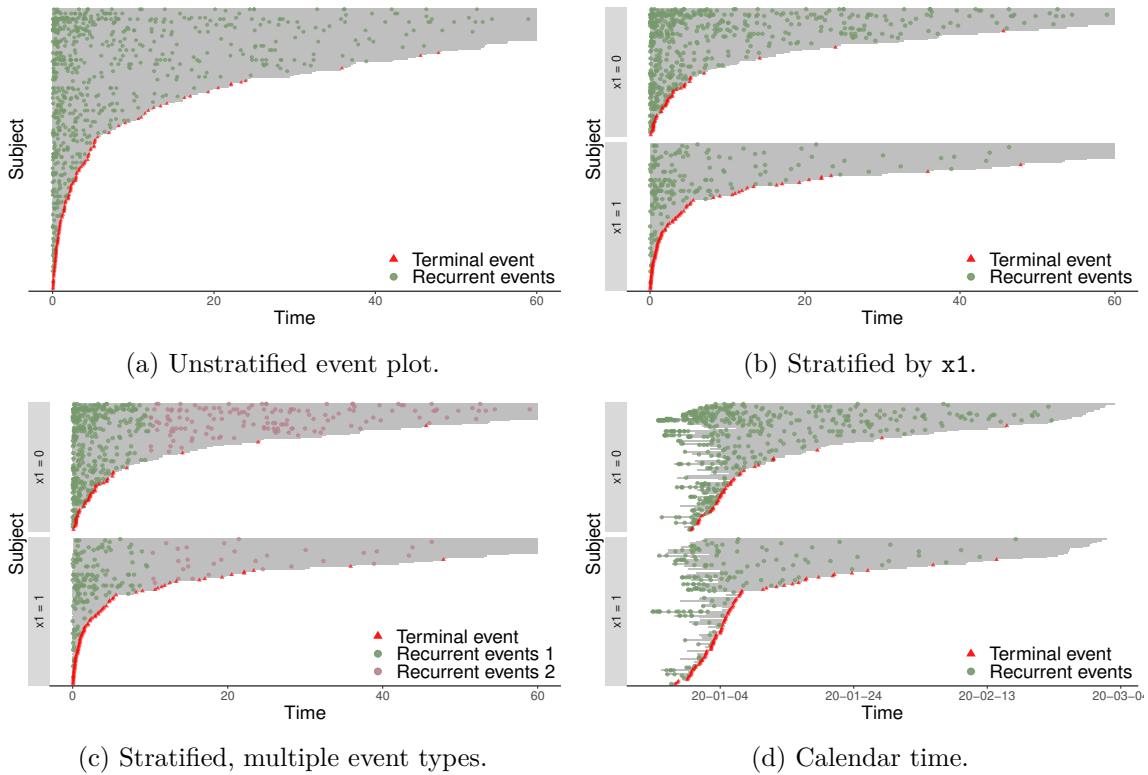

Figure 2: Event plots produced by `plot()` and `plotEvent()`.

```
R> args(basebind)

## function (..., legend.title, legend.labels, control = list())
## NULL
```

The argument "..." represents a list of `ggplot` objects created by plotting `reReg` objects. The arguments `legend.title` and `legend.labels` are optional character strings to control the legend title and legend labels in the combined plot, respectively. When fitting regression models with `reReg()`, the `baseline()` function can be applied to combine the estimates of baseline functions across groups. The following code is used to create Figure 3b, where the NPMLEs for `x1 = 0` and `x1 = 1` are presented. The corresponding 95% confidence intervals are computed with 200 bootstrap replicates.

```
R> mcf2 <- reReg(Recur(t.start %to% t.stop, id, event, status) ~ 1,
+    subset = x1 == 0, data = simDat, B = 200)
R> mcf3 <- update(mcf2, subset = x1 == 1)
R> g1 <- plot(mcf2)
R> g2 <- plot(mcf3)
```

```
R> basebind(g1, g2, legend.title = "X1", legend.labels = 0:1)
```

Under the independent censoring assumption, the **reReg** package imports the `mcf()` function from the **reda** package to compute the MCFs. The usage of the `mcf()` function is similar to that of the `plotEvent()`, as they both allows the recurrent event process to be specified in a model formula. When the overall sample MCF is of interest, the Nelson-Aalen estimate in Equation 1 can be created by applying the generic function `plot()` to the `Recur` object with an additional argument `mcf = TRUE`. This then internally calls the `mcf()` function with an intercept-only-model.

The following command plots the Nelson-Aalen estimate in Figure 3c, where the 95% confidence interval is enabled by additionally setting `mcf.conf.int = TRUE`.

```
R> plot(reObj, mcf = TRUE, mcf.conf.int = TRUE)
```

To create the cumulative sample mean function, one needs to additionally specify the argument `mcf.adjustRiskset = FALSE`. The stratified MCF can be visualized by first specifying the model formula with `mcf()`, then applying the generic function `plot()`. The following code is used to produce Figure 3d, which displays the Nelson-Aalen estimates for `x1 = 0` and `x1 = 1`.

```
R> mcf0 <- mcf(Recur(t.start %to% t.stop, id, event, status) ~ x1,
+    data = simDat)
R> plot(mcf0, conf.int = TRUE)
```

### 5.4. Fitting regression models

The main function for fitting semiparametric regression models in the **reReg** package is the `reReg()` function. When covariates are specified in the `reReg()` function, the `reReg()` function fits Model 6 using the following arguments

```
R> args(reReg)

## function (formula, data, subset, model = "cox", B = 0, se = c("boot",
##     "sand"), control = list())
## NULL
```

As in the `plotEvents()`, the arguments `formula` and `data` are used to specify the model formula and the optional data frame, respectively. The argument `model` is a character string used to specified the model type. The possible model types are `cox`, `ar`, `am`, and `gsc`, corresponding to the Cox-type model, the accelerated rate model, the accelerated mean model, and the generalized scale-change model, respectively. When the interest is the covariate effects on the risk of recurrent events and the terminal event is treated as nuisances, `model = "cox"` and `model = "gsc"` give the Cox-type model (4) and the generalized scale-change rate model considered in Wang *et al.* (2001) and Xu *et al.* (2020), respectively. When the recurrent event process and the terminal events are modeled jointly, the types of rate function and hazard function can be specified simultaneously in `model` separated by "`|`". The possible model types for the hazard function are `cox`, `ar`, `am`, and `gsc`. For examples, the joint frailty Cox-type model of Huang and Wang (2004) and the joint frailty accelerated mean model of Xu *et al.* (2017) can be called by `model = "cox|cox"` and `model = "am|am"`, respectively. Depending on the application, users can specify different model types for the rate function and the hazard function. For example, `model = "cox|ar"` postulate a Cox-type proportional model for the recurrent event rate function and an accelerated rate model for the terminal event hazard function, or $\alpha = \theta = 0$ in Model 6.

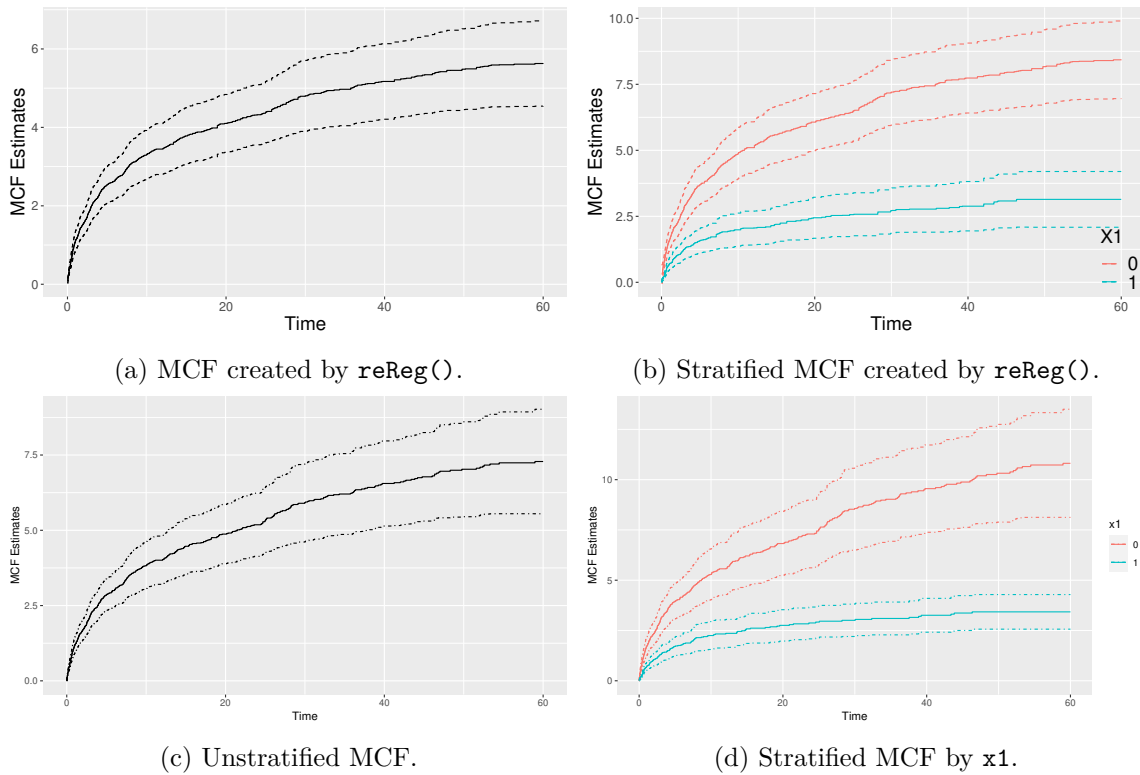

Figure 3: MCF plots produced by `plot()`.

For inference, the variance estimates are obtained from the nonparametric bootstrap approach with the argument `se = "boot"`. When fitting the generalized scale-change rate model of Xu *et al.* (2020), an efficient resampling-based sandwich estimator is also available via `se = "sand"`. The number of bootstrap or resampling is controlled by the argument B. When `B = 0` (default), the variance estimation procedure will not be performed and only the point estimates will be returned, which can be useful when the variance estimation is time consuming. The argument `control` is a list specifying changes to the algorithm control parameters. The full list of control parameters is:

```
R> args(reReg.control)

## function (eqType = c("logrank", "gehan", "gehan_s"), solver = c("BB::dfsane",
##     "BB::BBsolve", "BB::BBoptim", "optim"), tol = 1e-07, init = list(alpha = 0,
##     beta = 0, eta = 0, theta = 0), boot.parallel = FALSE, boot.parCl = NULL,
```

```
##     maxit1 = 100, maxit2 = 10, trace = FALSE, numAdj = 1e-07)
## NULL
```

The argument `eqType` is a character string used to specify the type estimating equation is used in the estimating procedure. The available options are the log-rank-type (`"logrank"`), the Gehan-type (`"gehan"`), or the induced smoothing Gehan-type (`"gehan_s"`) estimating equation is used in the estimating procedure. The default value is `"logrank"` corresponding to setting $\phi_i(\alpha) = 1$ in $S_{1n}(\alpha)$ and $\varphi_i(\eta, \theta) = 1$ in $S_{3n}(\eta, \theta)$ and $S_{4n}(\eta, \theta)$. The argument `solver` is a character string used to specify the equation solver used in solving the estimating equations. The default equation solver (`"BB::dfsane"`) uses the derivative-free Barzila-Borwein spectral approach for solving nonlinear equations implemented in `dfsane()` from the package **BB** (Varadhan and Gilbert 2009). Setting `solver = "BB::BBsolve"` calls the wrapper function `BBsolve()` in the **BB** package to locate a root with different Barzilai-Borwein steplengths, non-monotonicity parameters, and initialization approaches. Based on our observation, the `"BB::BBsolve"` algorithm generally exhibited more reliable convergence but the `solver = "BB::dfsane"` algorithm provides a better balance between convergence and speed. Alternative options are `solver = "BB::BBoptim"` and `solver = "optim"` that attempt to identify roots by minimizing the $\ell_2$-norm of the estimating functions. The options `solver = "BB::BBoptim"` and `solver = "optim"` call the `BBoptim()` function from the package **BB** and the base function `optim()`, respectively. The argument `tol` is the absolute tolerance used in the convergence criteria, while the arguments arguments `maxit1` and `maxit2` control the maximum numbers of iterations in the risk model and the hazard model, respectively. The argument `numAdj` specifies the $\epsilon$ used in the heuristic adjustment. The argument `init` is a list of initial values used in the root-finding algorithms. The list members `alpha`, `beta`, `eta`, and `theta` correspond to the parameters $\alpha$, $\beta$, $\eta$, and $\theta$ in Model 6, respectively. The default values for these initial values are zeros.

In an attempt to overcome the computational burden in bootstrap variance estimation, parallel computing techniques based on methods in the **parallel** package will be applied when `boot.parallel = TRUE`. The number of CPU cores used in the parallel computing is controlled by the argument `boot.parCl`, whose default value is half of the total number of CPU cores available on the current host identified by `parallel::detectCores() %/% 2`.

To illustrate the usage of the `reReg()` function, we first fit Cox-type rate model (4) to the `simDat` data with the variance estimate obtained from the nonparametric bootstrap approach with 200 bootstrap replicates in the following.

```
R> fm <- Recur(t.stop, id, event, status) ~ x1 + x2
R> set.seed(0)
R> system.time(fit1 <- reReg(fm, data = simDat, model = "cox", B = 200))

##    user  system elapsed
##   0.972   0.008   0.982
```

In this example, the implemented method finished in a reasonable computing time, which is based on a Linux machine with Core i7-6700@3.40 GHz processor and without parallel computing in the variance estimation. The summary of the model is presented below.

```
R> summary(fit1)

## Call:
## reReg(formula = fm, data = simDat, model = "cox", B = 200)
##
## Recurrent event process:
##    Estimate   StdErr z.value   p.value
## x1 -1.00483  0.16380 -6.1344 8.547e-10 ***
## x2 -0.97517  0.13305 -7.3292 2.316e-13 ***
```

The summary suggests statistically significant negative effects of both covariates on the rate function of the recurrent event process. This indicates that subjects with larger `x1` and `x2` values are likely to experience less frequently throughout the follow-up. In this case, we assumed the primary interest is the covariate effects on the risk of recurrent events and treat the terminal events as nuisances. When it is of interest to analyze the recurrent events and the terminal events simultaneously, we fit the joint Cox-type model (5) and present the summary as below:

```
R> set.seed(0)
R> system.time(fit2 <- reReg(fm, data = simDat, model = "cox|cox", B = 200))

##    user  system elapsed
##   1.276   0.032   1.315
```

```
R> summary(fit2)

## Call:
## reReg(formula = fm, data = simDat, model = "cox|cox", B = 200)
##
## Recurrent event process:
##    Estimate   StdErr z.value   p.value
## x1 -1.00483  0.16380 -6.1344 8.547e-10 ***
## x2 -0.97517  0.13305 -7.3292 2.316e-13 ***
##
## Terminal event:
##    Estimate  StdErr z.value   p.value
## x1  1.05295 0.28842  3.6507 0.0002615 ***
## x2  0.85086 0.26647  3.1931 0.0014073 **
```

The top panel of the summary is identical to that from fitting the Cox-type rate model (4) because the rate models in Models 4 and 5 are identical and the parameters therein are obtained from the same estimating equation. This reiterates the implemented estimation procedure is in a two-step fashion. When the terminal events are modeled as in `fit2`, `reReg()` additionally compute the parameters in the hazard model with the results presented in the lower panel of the summary. The summary exhibits statistically significant positive covariate effects on the hazard function, indicating that subjects with larger `x1` and `x2` values are in higher risk of terminal events.

The estimates of the baseline cumulative functions, e.g., $\Lambda_0(\cdot)$ and $H_0(\cdot)$, can be plotted by applying the generic function `plot()` to the `reReg` object. The following shows the S3 method for plotting a `reReg` object.

```
R> argsAnywhere(plot.reReg)

## function (x, baseline = c("both", "rate", "hazard"), smooth = FALSE,
##     newdata = NULL, frailty = NULL, showName = FALSE, control = list(),
##     ...)
## NULL
```

When a joint model is specified in `reReg()`, the `baseline` argument provides options to display only the baseline cumulative rate function with `baseline = "rate"`, only the baseline cumulative hazard function with `baseline = "hazard"`, or both the baseline cumulative rate function and the baseline cumulative hazard function with `baseline = "both"` (default). Even though the identifiability assumption, $\Lambda_0(\tau) = 1$, is used in the estimating procedure, we scaled $\widehat\Lambda_{0n}(\cdot)$ by $\mu_Z$ so the values represent the expected number of recurrent events. The `smooth` argument is a logical variable specifying whether the baseline cumulative functions will be smoothed by the monotonic increasing *P*-spline (Pya and Wood 2015) implemented in the **scam** package (Pya 2020).

The `plot()` method also allows user to plot the predicted cumulative functions given covariates. As in many prediction models, the plot method allows users to specify the covariates in a data frame via the argument `newdata`. When `newdata` is not specified (or is `NULL`), `newdata` is set to zeros and the baseline cumulative functions are produced. When `newdata` has more than one rows, each unique row will be treated as an observation and the corresponding predicted cumulative functions will be plotted. The `frailty` argument specifies the frailty value to be used in constructing the predicted cumulative functions; $\widehat\mu_Z$ is used when `frailty` is `NULL`. The `showName` is a logical variable indicating whether to label the objects' name at the end of the curves. The objects' names are determined by the row names of the `newdata`. The `control` is an optional list containing graphical parameters including `xlab`, `ylab`, `main`, etc. As the plot method for `Recur` objects, a `ggplot2` object is returned to allow easy customization.

The baseline cumulative rate function and the baseline cumulative hazard function of the joint frailty Cox model is plotted below and presented in Figure 4a.

```
R> plot(fit2)
```

To illustrate the feature in predicting cumulative functions given covariates, we consider two hypothetical subjects, one with `x1 = 0` and the other with `x1 = 1`. Holding the value of `x2` at the average, the predicted cumulative functions are presented in Figures 4b using the following codes

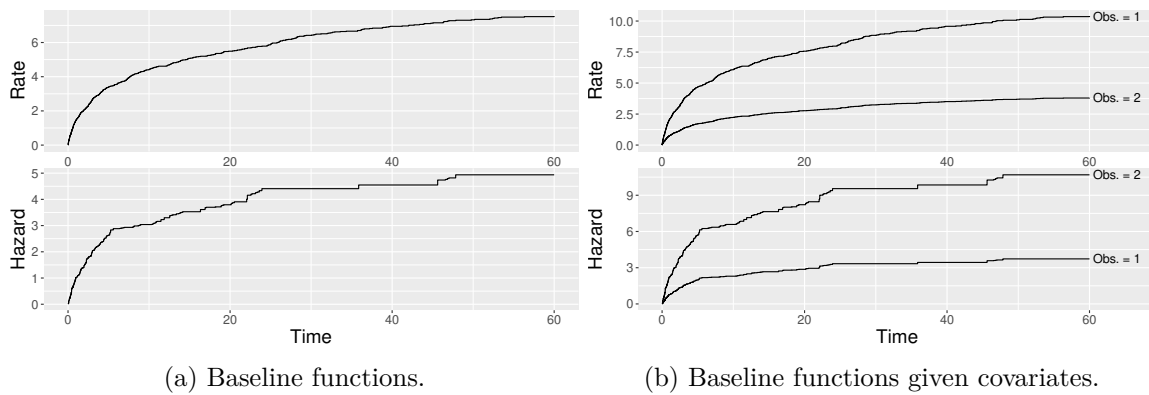

(a) Baseline functions.
(b) Baseline functions given covariates.

Figure 4: Plots of baseline functions produced by `plot()` and `plotRate()`.

```
R> newdata <- expand.grid(x1 = 0:1, x2 = mean(simDat$x2))
R> plot(fit2, newdata = newdata, showName = TRUE)
```

In practice, the generalized scale-change model is appealing when the actual form of the rate function is unknown. The following command fits the generalized scale-change model of Xu *et al.* (2020) by specifying `model = "gsc"`. The resampling-based sandwich estimator with 200 bootstrap replicates is used to obtain the standard error estimation at the default settings.

```
R> set.seed(0)
R> system.time(
+    fit3 <- reReg(fm, data = simDat, model = "gsc", se = "sand", B = 200))

##    user  system elapsed
##   1.408   0.016   1.424

R> summary(fit3, test = TRUE)

## Call:
## reReg(formula = fm, data = simDat, model = "gsc", B = 200, se = "sand")
##
## Recurrent event process (shape):
##     Estimate    StdErr z.value p.value
## x1 -0.022521  0.373706 -0.0603  0.9519
## x2 -0.114796  0.287355 -0.3995  0.6895
##
## Recurrent event process (size):
##     Estimate   StdErr z.value   p.value
## x1 -1.01380  0.26309 -3.8534 0.0001165 ***
## x2 -1.04078  0.23562 -4.4172 9.997e-06 ***
##
## Hypothesis tests:
## Ho: shape = 0 (Cox-type model):
```

```
##      X-squared = 0.1653, df = 2, p-value = 0.9207
## Ho: size = 0 (Accelerated rate model):
##      X-squared = 29.7184, df = 2, p-value = 0
## Ho: shape = size (Accelerated mean model):
##      X-squared = 43.5266, df = 2, p-value = 0
```

The summary statistics for the shape parameters and the size parameters are presented in different panels. When the additional option of `test = TRUE` is specified in `summary()` as in the above example, the hypothesis testing approach for submodel selection proposed in Xu *et al.* (2020) is performed. The null hypotheses are $H_o : \alpha = 0$, $H_o : \alpha = \beta$, and $H_o : \beta = 0$, corresponding to the Cox-type model, the accelerated rate model, and the accelerated mean model, respectively. The above results indicates a strong evidence that the rate function has a Cox structure.

### 5.5. Other common recurrent event models

Under the non-informative censoring assumption, the AG model with robust variance estimation can be called via `reReg()` with `model = "cox.LWYY"`:

```
R> summary(reReg(fm, data = simDat, model = "cox.LWYY"))

## Call:
## reReg(formula = fm, data = simDat, model = "cox.LWYY")
##
## Fitted with the Cox model of Lin et al. (2000):
## Recurrent event process:
##    Estimate   StdErr z.value   p.value
## x1 -1.13602  0.13704 -8.2900 < 2.2e-16 ***
## x2 -1.07493  0.14264 -7.5361 4.841e-14 ***
```

This is equivalent to calling the `coxph()` from the **survival** package with the `cluster` option:

```
R> library(survival)
R> summary(coxph(Surv(t.start, t.stop, event) ~ x1 + x2 + cluster(id),
+    data = simDat))$coef

##          coef exp(coef)  se(coef) robust se        z      Pr(>|z|)
## x1 -1.136023 0.3210936 0.08866522 0.1370350 -8.290017 1.132320e-16
## x2 -1.074935 0.3413201 0.07478383 0.1426375 -7.536128 4.841298e-14
```

The joint frailty scale-change model (6) also includes many models as special cases under the non-informative censoring assumption when the frailty is degenerated, i.e., $Z = 1$. Some of these special cases include the Cox-type models of Pepe and Cai (1993); Lin *et al.* (2000); Ghosh and Lin (2002), accelerated rate model Chen and Wang (2000); Ghosh (2004), accelerated means regression Ghosh and Lin (2003), and a general class of accelerated means model Sun and Su (2008). Different estimating procedures are proposed in each method. Of these, setting `model = "cox.LWYY"`, `model = "cox.GL"`, and `model = "am.GL"`, calls the methods proposed by Lin *et al.* (2000), Ghosh and Lin (2002), and Ghosh and Lin (2003), respectively.

### 5.6. Simulating recurrent event data

The **reReg** package allows users to generate simulated data from the generalized frailty joint scale-change model (6) via function `simGSC()`. The arguments of `simGSC()` are presented below.

```
R> args(simGSC)

## function (n, summary = FALSE, para, xmat, censoring, frailty,
##     tau, origin, Lam0, Haz0)
## NULL
```

The only required argument is the number of observation, represented by `n`. The remaining arguments are optional and will be assigned default values when not specified. The argument `summary` is a logical value indicating whether descriptive statistics of the simulated data will be printed after the data generation. The argument `para` is a list of regression parameters. The list members `alpha`, `beta`, `eta`, and `theta` are $p$-dimensional numerical vectors corresponding to regression parameters, $\alpha, \beta, \eta$, and $\theta$ in Model 6, respectively. The `xmat` argument is the $n \times p$ design matrix. The `censoring` and `frailty` arguments are $n$-dimensional numerical vectors that specify the independent censoring time, $C$, and the frailty variable, $Z$, respectively. The argument `tau` is the maximum follow-up time $\tau$ and the argument `origin` the time origin for each subject. Finally, the arguments `Lam0` and `Haz0` are single argument functions used to specify the baseline cumulative rate function, $\Lambda_0(t)$, and the baseline cumulative hazard function, $H_0(t)$, respectively.

At the default setting, the `simGSC()` function assumes $p = 2$ and the regression parameters to be $\alpha = \eta = (0,0)^\top$, $\beta = (-1,-1)^\top$, and $\theta = (1,1)^\top$. When the `xmat` and the `censoring` arguments are not specified, the `simGSC()` function assumes $X_i$ is a two-dimensional vector $X_i = (X_{i1}, X_{i2}), i = 1, \ldots, n$, where $X_{i1}$ is a Bernoulli variable with rate 0.5 and $X_{i2}$ is a standard normal variable. With the default `xmat`, the censoring time $C$ is generated from an independent uniform distribution in $[0, 2\tau X_{i1} + 2Z^2\tau(1 - X_{i1})]$. Thus, the censoring distribution is covariate dependent and is informative when $Z$ is not a constant. On the other hand, when `xmat` is specified by the users, the censoring time $C$ is generated from an independent uniform distribution $[0, 2\tau]$. When the `frailty` argument is not specified, the frailty variable $Z$ is generated from a gamma distribution with a unit mean and a variance of 0.25. The default values for `tau` and `origin` are 60 and 0, respectively. When arguments `Lam0` and `Haz0` are left unspecified, the `simGSC()` function uses $\Lambda_0(t) = 2\log(1+t)$ and $H_0(t) = \log(1+t)/5$, respectively. This is equivalent to setting `Lam0 = function(x) 2 * log(1 + x)` and `Haz0 = function(x) log(1 + x) / 5`. In summary, the default specifications generate the recurrent events and the terminal events from the model:

$$\begin{cases} \lambda(t) = \dfrac{2Z}{1 + te^{-X_{i1} - X_{i2}}}, \\ h(t) = \dfrac{Z}{5(1 + te^{X_{i1} + X_{i2}})}, \quad t \in [0, 60]. \end{cases}$$

The `simGSC()` function generates simulated data from the above specification and returns a `data.frame` in the same format as the built-in data set, `simDat`. Specifically, `simDat` was generated using the default settings of `simGSC()`.

```
R> data("simDat")
R> set.seed(0); dat <- simGSC(200, summary = TRUE)

## Call:
## simGSC(n = 200, summary = TRUE)
##
## Summary:
## Sample size:                                      200
## Number of recurrent event observed:               674
## Average number of recurrent event per subject:    3.37
## Proportion of subjects with a terminal event:     0.59
## Median time-to-terminal event:                    6.975

R> identical(dat, simDat)

## [1] TRUE
```

The following example illustrates the computational performance of the implemented method. We simulated data from `simGSC()` under the default settings. Each data is then fitted with the joint Cox-type model (Huang and Wang 2004), the joint accelerated mean model (Xu *et al.* 2017), and the general scale-change model (Xu *et al.* 2020), corresponding to `model = "cox|cox"`, `model = "am|am"`, and `model = "gsc"`, respectively. The reported computing times are without variance estimation and are averaged over 50 Monte Carlo runs.

```
R> fm <- Recur(t.stop, id, event, status) ~ x1 + x2
R> sizes <- c(100, 200, 400, 600, 800, 1000)
R> set.seed(0)
R> times <- sapply(sizes, function(n) {
+    rowMeans(replicate(50, {
+      dat <- simGSC(n)
+      c(system.time(reReg(fm, data = dat, model = "cox|cox", B = 0))[3],
+        system.time(reReg(fm, data = dat, model = "am|am", B = 0))[3],
+        system.time(reReg(fm, data = dat, model = "gsc", B = 0))[3])
+    }))
+ })
```

The following summarizes the average computing times in Table 2.

```
R> rownames(times) <-
+    c("\\texttt{model = cox|cox}", "\\texttt{model = am|am}",
+      "\\texttt{model = gsc}")
R> kable(round(times, 3), col.names = paste0("n = ", sizes),
```

```
+     caption = "Average runtime (in seconds) under different sample sizes.",
+     row.names = TRUE, booktabs = TRUE, escape = FALSE)
```

|                  | n = 100 | n = 200 | n = 400 | n = 600 | n = 800 | n = 1000 |
|------------------|---------|---------|---------|---------|---------|----------|
| `model = cox|cox` | 0.015   | 0.018   | 0.027   | 0.038   | 0.052   | 0.070    |
| `model = am|am`   | 0.101   | 0.189   | 0.381   | 0.511   | 0.769   | 1.112    |
| `model = gsc`     | 0.053   | 0.083   | 0.134   | 0.207   | 0.254   | 0.320    |

Table 2: Average runtime (in seconds) under different sample sizes.

The joint Cox-type model is the fastest among the three models for all sample sizes considered. Other than the sample size, several factors could impact the computing speed, such as the number of recurrent events and covariates. A comprehensive illustration of the computational performance of the implemented method under different scenarios might worth further investigation. Nonetheless, Table 2 suggests that analyses can be made available within seconds or minutes using the **reReg** package.

## 6. Application to colorectal cancer data

We demonstrate the usage of the **reReg** package to the `colorectal` dataset from the **frailtypack** package (Rondeau *et al.* 2012). The dataset consists of a random selection of 150 patients with advanced colorectal cancer enrolled in a randomized phase III clinical trial FFCD 2000-05 conducted between 2002 and 2007 (Ducreux, Malka, Mendiboure, Etienne, Texereau, Auby, Rougier, Gasmi, Castaing, Abbas, Pierre, Gargot, Azzedine, Lombard-Bohas, Geoffroy, Denis, Pignon, Bedenne, and Bouché 2011). The enrolled patients were randomized to receive either sequential or combination chemotherapy. One of the primary interests is to compare the two treatment groups in disease progression and overall survival. In this study, the recurrent events are identified as the appearances of new lesions, and the terminal event is death. The structure of the `colorectal` dataset is presented below.

```
R> data("colorectal", package = "frailtypack")
R> str(colorectal, vec.len = 2)

## 'data.frame': 289 obs. of  10 variables:
##  $ id            : int  1 2 3 3 3 ...
##  $ time0         : num  0 0 ...
##  $ time1         : num  0.71 1.28 ...
##  $ new.lesions   : int  0 0 1 1 0 ...
##  $ treatment     : Factor w/ 2 levels "S","C": 1 2 1 1 1 ...
##  $ age           : Factor w/ 3 levels "<60 years","60-69 years",..: 2 3 2 2 2 ...
##  $ who.PS        : Factor w/ 3 levels "0","1","2": 1 1 2 2 2 ...
##  $ prev.resection: Factor w/ 2 levels "No","Yes": 1 1 1 1 1 ...
##  $ state         : int  1 1 0 0 1 ...
##  $ gap.time      : num  0.71 1.28 ...
```

There are 150 sampled patients with 77 in the sequential treatment group. Key variables needed to create a recurrent event object with `Recur()` are subject identification (`id`), start and end of time intervals (`time0`, `time1`), recurrent event indicator (`new.lesions`), and terminal event indicator (`state`). The recurrent event object and its summary are printed with the following codes.

```
R> (colreObj <- with(colorectal,
+    Recur(time0 %to% time1, id, new.lesions, state)))

## [1] 1: (0.0000, 0.7096*]
## [2] 2: (0.0000, 1.2822*]
## [3] 3: (0.0000, 0.5246], (0.5246, 0.9208], (0.9208, 0.9425*]
## [4] 4: (0.0000, 0.6639], (0.6639, 0.7178*]
## [5] 5: (0.0000, 0.1585], (0.1585, 0.3689], (0.3689, 0.4630*]
##  [ reached getOption("max.print") -- omitted 145 entries ]

R> summary(colreObj)

## Call:
## Recur(time = time0 %to% time1, id = id, event = new.lesions,
##     terminal = state)
##
## Sample size:                                     150
## Number of recurrent event observed:              139
## Average number of recurrent event per subject:   0.927
## Proportion of subjects with a terminal event:    0.807
## Median follow-up time:                           1.199
## Median time-to-terminal event:                   1.351
```

From the descriptive statistics, the average number of new lesions recorded per patient is 0.927. Among the 150 patients, the proportion of subjects with a terminal event is 80.7%, the median follow-up time is 1.20 years, and the median time to death is 1.35 years. The recurrent event history of the patients in both treatment groups are displayed in Figure 5a, where the event plot is created by the `plotEvents()` function in the following.

```
R> fm <- Recur(time0 %to% time1, id, new.lesions, state) ~ treatment
R> plotEvents(fm, data = colorectal,
+    recurrent.name = "New lesions", terminal.name = "Death")
```

Additional graphical arguments are included to improve the readability of the event plot. Overall, patients in the sequential treatment group seem to have more new lesions early, while the number of deaths appears to spread out. On the other hand, the MCF estimates presented in Figure 5b suggest that the MCF estimates for the two treatment groups are not significantly different, as indicated by the overlapping 95% confidence intervals.

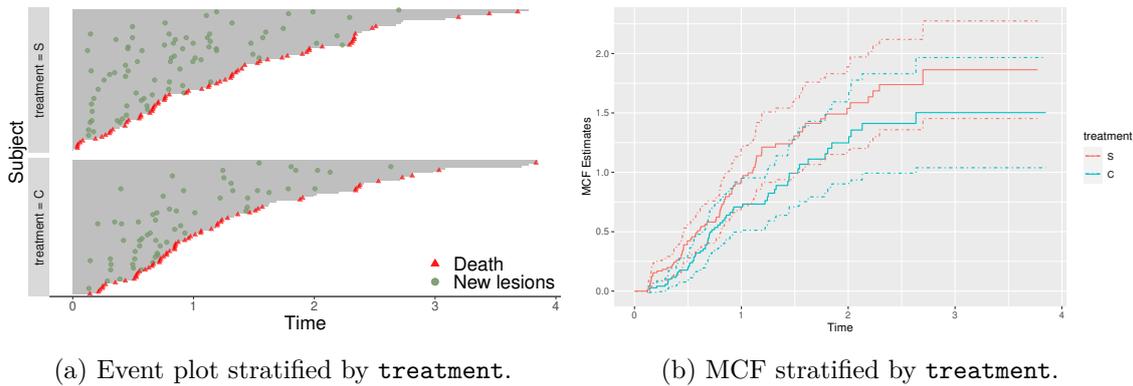

(a) Event plot stratified by `treatment`.  (b) MCF stratified by `treatment`.

Figure 5: Event plot and MCF plot stratified by treatment type.

```
R> plot(mcf(fm, data = colorectal), conf.int = T)
```

To evaluate the treatment effect in regression models, we first fit the joint Cox-type model (5) using the sequential treatment group as the reference group. In addition to the treatment effect, we also include baseline covariates including age (`age`), World Health Organization (WHO) performance status (`who.PS`), and previous resection indicator (`prev.resection`). The age variable is categorized into three groups; $< 60$ years, 60–69 years, and $> 69$ years. The WHO performance status is used to quantify patients' ability to carry out daily tasks. The available WHO performance status scores are 0, 1, and 2, where fully active patients who can carry out all activities without restriction have a score of 0. The following results show that, though not statistically significant, patients in the combination treatment group were more likely to experience fewer new lesions ($\hat{\beta} = -0.240, p = 0.436$) and had a lower risk of death ($\hat{\theta} = -0.088, p = 0.810$) than those in the sequential treatment group.

```
R> set.seed(0)
R> fm <- update(fm, ~ .+ age + who.PS + prev.resection)
R> fitCox <- reReg(fm, data = colorectal, model = "cox|cox", B = 200)
R> summary(fitCox)

## Call:
## reReg(formula = fm, data = colorectal, model = "cox|cox", B = 200)
##
## Recurrent event process:
##                     Estimate    StdErr z.value p.value
## treatmentC         -0.240316  0.308706 -0.7785  0.4363
## age60-69 years     -0.368456  0.347377 -1.0607  0.2888
## age>69 years       -0.277112  0.383778 -0.7221  0.4703
## who.PS1            -0.323627  0.349054 -0.9272  0.3538
## who.PS2             0.084318  0.353789  0.2383  0.8116
## prev.resectionYes  -0.240201  0.282869 -0.8492  0.3958
##
## Terminal event:
```

```
##                       Estimate    StdErr z.value p.value
## treatmentC           -0.087141  0.366157 -0.2380  0.8119
## age60-69 years       -0.259770  0.351031 -0.7400  0.4593
## age>69 years         -0.346661  0.434668 -0.7975  0.4251
## who.PS1              -0.322803  0.433917 -0.7439  0.4569
## who.PS2               0.523742  0.404880  1.2936  0.1958
## prev.resectionYes    -0.453941  0.362956 -1.2507  0.2111
```

We next fit a generalized scale-change model in the following to gain more insights into the treatment effects on the recurrent event progress. The results suggest that patients in the combination treatment group tended to have a decelerated time to new lesions by a factor of $1/\exp(-0.969) = 2.635, p = 0.009$. This implies patients in the combination treatment group were more likely to experience new lesions later. The hypothesis testing results suggest an accelerated rate model or an accelerated mean model on the recurrent event process.

```
R> set.seed(0)
R> fitGSC <- reReg(fm, data = colorectal, model = "gsc", se = "sand", B = 200)
R> summary(fitGSC, test = TRUE)

## Call:
## reReg(formula = fm, data = colorectal, model = "gsc", B = 200,
##     se = "sand")
##
## Recurrent event process (shape):
##                       Estimate     StdErr z.value  p.value
## treatmentC           -0.9687721  0.3682715 -2.6306 0.008524 **
## age60-69 years        0.0041879  0.3333505  0.0126 0.989976
## age>69 years         -0.2704151  0.2640631 -1.0241 0.305809
## who.PS1              -0.2765749  0.2834604 -0.9757 0.329209
## who.PS2              -0.1967113  0.3369671 -0.5838 0.559375
## prev.resectionYes    -0.5218806  0.5238343 -0.9963 0.319119
##
## Recurrent event process (size):
##                       Estimate    StdErr z.value p.value
## treatmentC            0.148253  0.515998  0.2873  0.7739
## age60-69 years       -0.484778  0.488547 -0.9923  0.3211
## age>69 years         -0.239504  0.351567 -0.6812  0.4957
## who.PS1              -0.363964  0.578832 -0.6288  0.5295
## who.PS2               0.080312  0.393328  0.2042  0.8382
## prev.resectionYes     0.033565  0.521850  0.0643  0.9487
##
## Hypothesis tests:
## Ho: shape = 0 (Cox-type model):
##      X-squared = 17.3385, df = 6, p-value = 0.0081
## Ho: size = 0 (Accelerated rate model):
##      X-squared = 3.6693, df = 6, p-value = 0.7213
```

```
## Ho: shape = size (Accelerated mean model):
##      X-squared = 4.3678, df = 6, p-value = 0.627
```

# 7. Discussion

The **reReg** package provides a comprehensive toolkit for analyzing recurrent event data. It allows easy access to create event plots and MCF plots for exploratory data analysis and a generalized joint frailty scale-change model for regression analysis. The implemented methods accommodate informative censoring via the use of a subject-specific frailty variable. In contrast to existing frailty models, the implemented estimation procedure does not require distributional information on the frailty variable. Using the borrowing-strength approach in the estimating procedure, our model allows users to specify any combination of the sub-models between the recurrent event process and the terminal events when they are fitted jointly. Since the **reReg** package's debut on CRAN, it has been applied in many medical studies (e.g. Richter, Sierocinski, Singer, Bülow, Hackmann, Chenot, and Schmidt 2020; Ejoku, Odhiambo, and Chaba 2020; Deo, Sundaram, Sahadevan, Selvaganesan, Mohan, Rubelowsky, Josephson, Elgudin, Kilic, and Cmolik 2021).

Future work will be devoted to improving computational efficiency. In particular, we plan to generalize the efficient resampling-based sandwich estimator to all the sub-models. In addition, we plan to apply the induced smoothing method (e.g., Brown and Wang 2007; Chiou, Kang, and Yan 2015) to facilitate numerical reliability. Including different types of recurrent event models such as the additive rate model and the semiparametric transformation model would be of interest too. When longitudinal outcomes at recurrent events are available, we plan to expand the current version of the package to adopt time-dependent covariates (e.g., Huang *et al.* 2010). Another interesting extension is to use multiple frailty variables to allow different frailty effects.

# Acknowledgments

This research was partially supported by National Science Foundation SES-1846747 and Institute of Education Sciences R305D200015 for Xu, and National Institute of Health R01CA193888 for Huang.

# References

Amorim LD, Cai J (2015). "Modelling Recurrent Events: A Tutorial for Analysis in Epidemiology." *International Journal of Epidemiology*, **44**(1), 324–333.

Andersen PK, Gill RD (1982). "Cox's Regression Model for Counting Processes: A Large Sample Study." *The Annals of Statistics*, **10**, 1100–1120.

Barzilai J, Borwein JM (1988). "Two-Point Step Size Gradient Methods." *IMA Journal of Numerical Analysis*, **8**(1), 141–148.

Brown BM, Wang YG (2007). "Induced Smoothing for Rank Regression with Censored Survival Times." *Statistics in Medicine*, **26**(4), 828–836.

Charles-Nelson A, Katsahian S, Schramm C (2019). "How to Analyze and Interpret Recurrent Events Data in the Presence of a Terminal Event: An Application on Readmission After Colorectal Cancer Surgery." *Statistics in Medicine*, **38**(18), 3476–3502.

Chen YQ, Jewell NP (2001). "On a General Class of Semiparametric Hazards Regression Models." *Biometrika*, **88**(3), 687–702.

Chen YQ, Wang MC (2000). "Estimating a Treatment Effect with the Accelerated Hazards Models." *Controlled Clinical Trials*, **21**(4), 369–380.

Chiou SH, Huang CY (2022). **reReg**: *Recurrent Event Regression*. R package version 1.4.4, URL http://github.com/stc04003/reReg.

Chiou SH, Kang S, Yan J (2015). "Rank-Based Estimating Equations with General Weight for Accelerated Failure Time Models: An Induced Smoothing Approach." *Statistics in Medicine*, **34**(9), 1495–1510.

Chiou SH, Xu G, Yan J, Huang CY (2018). "Semiparametric Estimation of the Accelerated Mean Model with Panel Count Data Under Informative Examination Times." *Biometrics*, **74**(3), 944–953.

Claggett B, Pocock S, Wei LJ, Pfeffer MA, McMurray JJV, Solomon SD (2018). "Comparison of Time-to-First Event and Recurrent-Event Methods in Randomized Clinical Trials." *Circulation*, **138**(6), 570–577.

Clement D (2013). **condGEE**: *Parameter Estimation in Conditional GEE for Recurrent Event Gap Times*. R package version 0.1-4, URL https://CRAN.R-project.org/package=condGEE.

Clement DY, Strawderman RL (2009). "Conditional GEE for Recurrent Event Gap Times." *Biostatistics*, **10**(3), 451–467.

Cook RJ, Lawless J (2007). *The Statistical Analysis of Recurrent Events*. New York: John Wiley & Sons, Wiley.

Cook RJ, Lawless JF, Nadeau C (1996). "Robust Tests for Treatment Comparisons Based on Recurrent Event Responses." *Biometrics*, **52**(2), 557–571.

Cox DR (1975). "Partial Likelihood." *Biometrika*, **62**(2), 269–276.

Deo SV, Sundaram V, Sahadevan J, Selvaganesan P, Mohan SM, Rubelowsky J, Josephson R, Elgudin Y, Kilic A, Cmolik B (2021). "Outcomes of Coronary Artery Bypass Grafting in Patients with Heart Failure with a Midrange Ejection Fraction." *The Journal of Thoracic and Cardiovascular Surgery*. doi:https://doi.org/10.1016/j.jtcvs.2021.01.035. In press.

Ducreux M, Malka D, Mendiboure J, Etienne PL, Texereau P, Auby D, Rougier P, Gasmi M, Castaing M, Abbas M, Pierre M, Gargot D, Azzedine A, Lombard-Bohas C, Geoffroy P, Denis B, Pignon JP, Bedenne L, Bouché O (2011). "Sequential Versus Combination

Chemotherapy for the Treatment of Advanced Colorectal Cancer (FFCD 2000–05): An Open-Label, Randomised, Phase 3 Trial." *The Lancet Oncology*, **12**(11), 1032–1044.

Efron B (1982). *The Jackknife, the Bootstrap and Other Resampling Plans*. SIAM.

Ejoku J, Odhiambo C, Chaba L (2020). "Analysis of Recurrent Events with Associated Informative Censoring: Application to HIV Data." *International Journal*, **9**, 21.

Fine JP, Yan J, Kosorok MR (2004). "Temporal Process Regression." *Biometrika*, **91**(3), 683–703.

Fu H, Luo J, Qu Y (2016). "Hypoglycemic Events Analysis via Recurrent Time-to-Event (HEART) Models." *Journal of Biopharmaceutical Statistics*, **26**(2), 280–298.

Ghosh D (2004). "Accelerated Rates Regression Models for Recurrent Failure Time Data." *Lifetime Data Analysis*, **10**(3), 247–261.

Ghosh D, Lin DY (2002). "Marginal Regression Models for Recurrent and Terminal Events." *Statistica Sinica*, **12**(3), 663–688.

Ghosh D, Lin DY (2003). "Semiparametric Analysis of Recurrent Events Data in the Presence of Dependent Censoring." *Biometrics*, **59**(4), 877–885.

Harrell Jr FE (2018). **rms**: *Regression Modeling Strategies*. R package version 5.1-2, URL https://CRAN.R-project.org/package=rms.

Huang CY, Qin J, Wang MC (2010). "Semiparametric Analysis for Recurrent Event Data with Time-Dependent Covariates and Informative Censoring." *Biometrics*, **66**(1), 39–49.

Huang CY, Wang MC (2004). "Joint Modeling and Estimation for Recurrent Event Processes and Failure Time Data." *Journal of the American Statistical Association*, **99**(468), 1153–1165.

Lawless JF, Nadeau C (1995). "Some Simple Robust Methods for the Analysis of Recurrent Events." *Technometrics*, **37**(2), 158–168.

Lin DY, Wei LJ, Yang I, Ying Z (2000). "Semiparametric Regression for the Mean and Rate Functions of Recurrent Events." *Journal of the Royal Statistical Society B*, **62**(4), 711–730.

Ma C, Qu Y, Fu H (2021). "Analysis of Recurrent Hypoglycemic Events." *Journal of Biopharmaceutical Statistics*, **31**(1), 5–13.

Pepe MS, Cai J (1993). "Some Graphical Displays and Marginal Regression Analyses for Recurrent Failure Times and Time Dependent Covariates." *Journal of the American Statistical Association*, **88**(423), 811–820.

Pya N (2020). **scam**: *Shape Constrained Additive Models*. R package version 1.2-8, URL https://CRAN.R-project.org/package=scam.

Pya N, Wood SN (2015). "Shape Constrained Additive Models." *Statistics and Computing*, **25**(3), 543–559.

Richter A, Sierocinski E, Singer S, Bülow R, Hackmann C, Chenot JF, Schmidt CO (2020). "The Effects of Incidental Findings from Whole-Body MRI on the Frequency of Biopsies and Detected Malignancies or Benign Conditions in a General Population Cohort Study." *European Journal of Epidemiology*, **35**(10), 925–935.

Rogers JK, Yaroshinsky A, Pocock SJ, Stokar D, Pogoda J (2016). "Analysis of Recurrent Events With an Associated Informative Dropout Time: Application of the Joint Frailty Model." *Statistics in Medicine*, **35**(13), 2195–2205.

Rondeau V, Mazroui Y, González JR (2012). "**frailtypack**: An R Package for the Analysis of Correlated Survival Data with Frailty Models Using Penalized Likelihood Estimation or Parametrical Estimation." *Journal of Statistical Software*, **47**(4), 1–28.

Sun L, Su B (2008). "A Class of Accelerated Means Regression Models for Recurrent Event Data." *Lifetime Data Analysis*, **14**(3), 357–375.

Therneau TM (2021). *A Package for Survival Analysis in R*. R package version 3.2-10, URL https://CRAN.R-project.org/package=survival.

Varadhan R, Gilbert P (2009). "**BB**: An R Package for Solving a Large System of Nonlinear Equations and for Optimizing a High-Dimensional Nonlinear Objective Function." *Journal of Statistical Software*, **32**(4), 1–26. URL http://www.jstatsoft.org/v32/i04/.

Wang MC, Chiang CT (2002). "Non-Parametric Methods for Recurrent Event Data with Informative and Non-Informative Censorings." *Statistics in Medicine*, **21**(3), 445–456.

Wang MC, Jewell NP, Tsai WY (1986). "Asymptotic Properties of the Product Limit Estimate Under Random Truncation." *The Annals of Statistics*, **14**, 1597–1605.

Wang MC, Qin J, Chiang CT (2001). "Analyzing Recurrent Event Data with Informative Censoring." *Journal of the American Statistical Association*, **96**(455), 1057–1065.

Wang W, Fu H, Chiou SH, Yan J (2021). **reda**: *Recurrent Event Data Analysis*. R package version 0.5.3, URL https://github.com/wenjie2wang/reda.

Wang X, Ma S, Yan J (2013). "Augmented Estimating Equations for Semiparametric Panel Count Regression with Informative Observation Times and Censoring Time." *Statistica Sinica*, **23**, 359–381.

Wickham H (2009). **ggplot2**: *Elegant Graphics for Data Analysis*. Springer-Verlag New York. ISBN 978-0-387-98140-6. URL http://ggplot2.org.

Xu G, Chiou SH, Huang CY, Wang MC, Yan J (2017). "Joint Scale-Change Models for Recurrent Events and Failure Time." *Journal of the American Statistical Association*, **112**(518), 794–805.

Xu G, Chiou SH, Yan J, Marr K, Huang CY (2020). "Generalized Scale-Change Models for Recurrent Event Processes Under Informative Censoring." *Statistica Sinica*, **30**, 1773–1795.

Yan J (2019). **tpr**: *Temporal Process Regression*. R package version 0.3-1.2, URL https://CRAN.R-project.org/package=tpr.

**Affiliation:**

Sy Han Chiou
Department of Mathematical Sciences
University of Texas at Dallas
800 W. Campbell Road,
Richardson, TX 75080, USA
E-mail: schiou@utdallas.edu

Gongjun Xu
Department of Statistics
University of Michigan
1085 South University Avenue,
Ann Arbor, MI 48109, USA
E-mail: gongjun@umich.edu

Jun Yan
Department of Statistics
University of Connecticut
215 Glenbrook Road U-4120,
Storrs, CT 06269, USA
E-mail: jun.yan@uconn.edu

Chiung-Yu Huang
Department of Epidemiology and Biostatistics
University of California, San Francisco
550 16th. Street,
San Francisco CA 94158, USA
E-mail: ChiungYu.Huang@ucsf.edu